\documentclass[aps,nofootinbib,superscriptaddress, showpacs,preprintnumbers,  twocolumn, nofootinbibt]{revtex4-1}
\usepackage{amsmath}
\usepackage{bm}
\usepackage{amsfonts}
\usepackage{amssymb}
\usepackage{mathrsfs}
\usepackage{color}
\usepackage{verbatim}
\usepackage{mwe}
\usepackage{graphicx}
\usepackage{makeidx}
\usepackage{eurosym}
\usepackage{tensor}
\usepackage{color}
\usepackage{revsymb}
\usepackage{hyperref}\hypersetup{
    colorlinks = true,
    linkcolor = blue,
    anchorcolor = blue,
    citecolor = blue,
    filecolor = blue,
    urlcolor = blue
   }
\def\be{\begin{equation}}
\def\ee{\end{equation}}
\def\bea{\begin{eqnarray}}
\def\eea{\end{eqnarray}}

\def\nn{\nonumber \\}

\def\nn{\nonumber}

\begin{document}

\title{Mimetic Weyl geometric gravity}
\author{Daria-Ioana Vișa}
\email{dariavisa29@gmail.com}
\affiliation{Department of Physics, Babes-Bolyai University, Kogalniceanu Street,
	Cluj-Napoca 400084, Romania,}
\author{Tiberiu Harko}
\email{tiberiu.harko@aira.astro.ro}
%\affiliation{Department of Theoretical Physics, National Institute of Physics
%and Nuclear Engineering (IFIN-HH), Bucharest, 077125 Romania,}
\affiliation{Department of Physics, Babes-Bolyai University, Kogalniceanu Street,
	Cluj-Napoca 400084, Romania,}
\affiliation{Astronomical Observatory, 19 Ciresilor Street,
	Cluj-Napoca 400487, Romania}
\author{Shahab Shahidi}
\email{s.shahidi@du.ac.ir}
\affiliation{School of Physics, Damghan University, Damghan, 41167-36716, Iran}
\date{\today }

\begin{abstract}
We consider a mimetic type extension of the Weyl geometric gravity theory, by assuming that the metric of the space-time manifold can be parameterized in terms of a scalar field, called the mimetic field. The action of the model is obtained by starting from a conformally invariant gravitational action, constructed, in Weyl geometry, from the square of the Weyl scalar, the strength of the Weyl vector, and an effective  matter term, respectively. After linearizing the action in the Weyl scalar by introducing an auxiliary scalar field, we include the mimetic field, constructed from the same auxiliary scalar field used to linearize the action, via a Lagrange multiplier. The conformal invariance of the action is maintained by imposing  the trace condition on the effective matter energy-momentum tensor, built up from the ordinary matter Lagrangian, and some specific functions of the Weyl vector, and the scalar field, respectively, thus making the matter sector of the action conformally invariant. The field equations are derived by varying the action with respect to the metric tensor, the Weyl vector field, and the scalar field, respectively. We investigate the cosmological implications of the Mimetic Weyl geometric gravity by considering the dynamics of  an isotropic and homogeneous Friedmann–Lemaitre–Robertson–Walker FLRW Universe.  The generalized Friedmann equations of the model are derived, and their solutions are obtained numerically for a dust filled Universe. Moreover, we perform a detailed comparison of the predictions of the considered model with a set of observational data for the Hubble function, and with the results of the $\Lambda$CDM standard paradigm. Our results indicate that the present model give a good description of the observational data, and they reproduce almost exactly the predictions of the $\Lambda$CDM scenario. Hence, Mimetic Weyl geometric gravity can be considered a viable alternative to the standard approaches to cosmology, and to the gravitational phenomena.
\end{abstract}

\maketitle
{
  \hypersetup{linkcolor=blue}
  \tableofcontents
}

\section{Introduction}

%{\bf Please add (present only ideas, without technical details; some simple mathematics may also be used):

%- brief discussion of the present-day situation in cosmology; mention the Hubble tensions

%-brief non-technical discussion of Weyl geometry, and its applications; discuss Weyl geometric gravity, kits origin, and applications

%-brief discussion of the mimetic gravity

%State the goal of the present work as follows (but feel free to develop the text):

%The goal of the present investigation is to extend the Weyl geometric gravity model by considering that the physical space-time
%metric is parameterized in terms of a conformal scalar degree of freedom, called mimetic field. }

The purpose of modern cosmology is to describe the evolution, origin, and large-scale structure of the Universe, seeking answers to fundamental questions through observational data, theoretical models and technological advancements. A huge advancement in our understanding of the Universe was the discovery of cosmic expansion, one of the most significant discoveries in cosmology. The first theoretical suggestions for the expansion of the Universe did appear in 1922, 1924 and 1927, respectively, in the works by Alexander Friedmann \cite{Fried1, Fried2} and by Georges Lema\^{i}tre \cite{Lemaitre1,Lemaitre2}, while the firm observational evidence for the global cosmological dynamics was found in 1929 by Edwin Hubble \cite{Hubble}. Hubble proved that there were galaxies beyond the Milky Way, and he did find the linear relationship between the radial velocities and distances from the study of the stars and mean luminosities of nebulae in a cluster. This relation is known as the Hubble law. The galaxies are moving away from us because the intergalactic space is expanding, and the expansion is detected via the redshift of the radiation traveling from the galactic sources to the Earth. These observations led to the formulation of the standard Big Bang cosmological theory, in which the Universe originated in a very special initial state, and expanded following the laws of general relativity.

A major change in the cosmological paradigm was represented by the discovery of the late accelerated expansion of the Universe \cite{E1,E2,E3}. One of the simplest explanations of the present day cosmic dynamics could be obtained by reintroducing the cosmological constant $\Lambda$ \cite{cc} into the gravitational field equations (note, however, that Einstein rejected this idea eventually \cite{ccrej}). The standard model of cosmology, called the $\Lambda$CDM (Lambda Cold Dark Matter), provides an exceptional fit to the cosmological data, yielding six free parameters and a set of well-chosen ansatzes.  Even if theoretically it could be described by the vacuum energy density obtained from quantum field theory considerations, the physical nature of $\Lambda$ remains elusive  (see reviews on the cosmological constant problem in \cite{cc0,cc0a,cc1,cc2}). To explain the observational results without resorting to the problematic cosmological constant, the concept of dark energy was introduced in cosmology (for reviews of the dark energy problem see \cite{rd1,rd2,rd3,rd4}. The cosmological constant is a particular form of dark energy, satisfying the equation of state $\rho_{DE}+p_{DE}=0$. There are three basic approaches to the dark energy problem \cite{rd5}, the dark constituents model, the dark gravity model, and the dark coupling theory. In the dark constituents model one modify the matter energy-momentum tensor by adding  additional degrees of freedom (typically scalar and dynamic alternatives for DE) such as, for example, the  quintessence field \cite{qui1,qui2,qui3}. In the dark gravity approach one assumes that dark energy is a modification of the gravitational force at large cosmological scales \cite{Od1,Od2}, while in the dark coupling theories one assumes the existence of a coupling between matter and geometry \cite{book}.  Detecting DE directly can be a challenging task, due to its lack of interaction with electromagnetic radiation. The XENON1T direct detection experiment reported detecting an unusual signal that could be explained by the absorption of dark energy scalars produced in the solar tachocline or any propagating scalar degree of freedom of a modified theory of gravity \cite{XEN}. To differentiate between DE as described by the cosmological constant, and the  modifications of GR, several experiments have been conducted, or are in the planning stage: DESI, which involves measuring baryon acoustic oscillations \cite{desi}, LSST \cite{lsst}, and Euclid \cite{euclid}, respectively.

The present rate of the expansion of the Universe is encapsulated in a single number, called the Hubble constant $H_0$ (or Hubble-Lema\^{i}tre constant), which sets the overall scale of the Universe. Determining the precise value of $H_0$ has become a major challenge in modern cosmology, due to the statistically significant tension between the late time and early time measurements of the present day value of the Hubble constant, with 4$\sigma$ to 6$\sigma$ disagreement \cite{t1,t2,t3, t4,t5}.
The Planck Cosmic Microwave Background team provided the most precise measurement of the Hubble constant $H_0 = 67.4 \pm 0.5 \rm{km s^{-1} Mpc^{-1}}$, with the best calibration of the cosmological parameters in $\Lambda$CDM model in 2021 \cite{CMB}. Other indirect measurements from the early Universe produce an identical outcome.
After multiple enhancements of the independent geometric calibrations of Cepheids, the SH0ES project and the Hubble Space Telescope (HST) provided the value of $H_0 = 73.30 \pm 1.04 \rm{km s^{-1} Mpc^{-1}}$ in 2022 \cite{CIA}.

The latest draft version \cite{jwst1} provided further evidence from the James Webb Space Telescope (JWST) that the systematic errors in the HST Cepheid photometry hold no significant relevance in the $\sim0.18$ mag Hubble Tension. JWST operates in the Near-infrared spectrum, providing more accurate and higher-quality results due to less interference from stellar crowding and cosmic dust \cite{jwst2}.

%Other independent methods that ensure consistent results with the previous methods are the baryon acoustic oscillations \cite{bao}, gravitational lensing \cite{gl}, %gravitational waves (standard sirens) \cite{gwbh1,gwbh2,gwbh3,gwbh4}, or as recently researched, through quasars \cite{q}. A more detailed description of recent Hubble %constant estimates is presented in \cite{hubbleten}.

Hence, it is important to note that the $\Lambda$CDM model fails to correctly describe some observations, as evidenced by the inconsistencies in $H_0$ measurements, where the values determined from the primordial epoch of the Universe yield a lower rate for the Hubble constant. If indeed it does exist, the Hubble tension points towards the necessity of considering some new cosmological and gravitational physics.

The second major problem present day cosmology faces is the problem of dark matter. The existence of Dark Matter was first evidenced in 1933 by Fritz Zwicky \cite{DM1}, where the luminous mass was determined to be only a small part of the mass required to prevent the galaxies from escaping the Coma Cluster. There have been numerous theories and experiments aimed at comprehending the nature of dark matter, yet it still remains unknown. Various models propose the existence of DM as hypothetical particles, including WIMPs \cite{Wimps1}, axions \cite{axion1,axion2,axion3}, sterile neutrinos \cite{wDMsn}, self-interacting DM \cite{sfdm1}, Bose-Einstein Condensates \cite{BH}, or as a modification of gravity, such as MOND \cite{MOND}, or modified gravity theories \cite{BH1}. Regarding the detection of DM,  the cluster merger 1E 0657-558, named the Bullet Cluster, enabled a direct detection for the non-luminous component of the Universe, revealing that the majority of the mass in the clusters is non-baryonic \cite{DM2}. %Other direct detection methods include Annual modulation signature(DAMA/LIBRA\cite{dama}, ANAIS-112\cite{anais} using NaI crystals), liquid Xenon %detectors (XENONnT \cite{xenon} , PandaX-4T\cite{panda},LUX-ZEPLIN\cite{lux}).

%When conducting indirect detection measurements, we examine particles produced(annihilation or decay) in various regions such as the Galactic Center, subhaloes, dwarf %spheroidals, the Sun via satellites or balloons, Cherenkov telescopes, and large neutrino observatories. Various states of matter can be measured, such as charged %particles(PAMELA\cite{pamela}, AMS \cite{ams1,ams2}), gamma(Cherenkov\cite{Cherenkov}, H.E.S.S.\cite{hess}, Fermi-LAT\cite{fermilat1,fermilat2}), %neutrinos(ANTARES\cite{Antares}, AMANDA-II and IceCube\cite{AMic}, IceCube-DeepCore\cite{icDC1,icDC2}, Super-Kamiokande\cite{k}, VERITAS\cite{veritas1,veritas2}). %Additionally, the LHC has provided constraints on the behaviour of dark matter particles in \cite{atlas1,atlas2,cms1,cms2}.

Understanding the geometry of spacetime is crucial for comprehending the nature of the Universe. Einstein \cite{einstein}, and Hilbert \cite{Hilbert} derived the equations of GR based on the Riemannian geometry. However, very soon after the development of general relativity,  Weyl proposed an extension of the mathematical framework of the gravitational theories, the Weyl geometry \cite{Weyl1,Weyl2,Weyl3}. For the historical development and physical applications of Weyl geometry see \cite{Scholz}.

 In GR, the transport of vector lengths is integrable. However, in Weyl's geometry, parallel transport takes into account the local properties of spacetime, resulting in variations in the relative size of vectors, when transported along curves. Thus, the parallel displacement of lengths is non-integrable.

Weyl generalized the  Riemannian geometry by introducing a new geometrical degree of freedom, the Weyl vector $\omega_\alpha$. In Weyl geometry the covariant derivative of the metric tensor does vanish, leading to the concept of non-metricity $Q_{\lambda\mu\nu}$, defined according to $\nabla _{\lambda} g_{\mu\nu} = Q_{\lambda\mu\nu}$, a relation known as the Weyl compatibility condition.

%Weyl's original attempt failed to unify the two fundamental forces due to physical inconsistencies as pointed out by Einstein \cite{Weyl2}, however, new developments %arised, leading to the emergence of gauge theories. In the present day, some important concepts were adopted.
Another important implication of Weyl’s geometry, is the concept of conformal invariance, which must be satisfied by all physical laws. This requires that the physical laws remain unchanged under local conformal transformations of the type $d\tilde{s}^2=\Sigma^n(x)ds^2=\Sigma^n(x)g_{\mu\nu}dx^{\mu}dx^{\nu}=\tilde{g}_{\mu\nu}dx^{\mu}dx^{\nu}$,  which relate to variations in the units of length and time at each point in the four-dimensional manifold. $\Sigma(x)$ is called the conformal factor, and $n$ is the Weyl charge.

Gravitational theories that fully implement the principle of conformal invariance are known as conformal gravity \cite{c1,c2}, and they are based on the action
\begin{equation}
I_W = -\alpha_g \int d^4x\sqrt{-g} C_{\lambda\mu\nu\kappa}C^{\lambda\mu\nu\kappa}
\end{equation}
where $C_{\lambda\mu\nu\kappa}$ is the Weyl conformal tensor,  and $\alpha_g$ the gravitational coupling constant.
This action is invariant under the local transformation $g_{\mu\nu}(x)\rightarrow e^{2\alpha(x)}g_{\mu\nu}(x)$. This theory can explain the behavior of the galactic rotation curves without the need for DM.

A generalization of Weyl's theory was proposed by Dirac \cite{Di1,Di2},  who extended the gravitational action by introducing  a real scalar field $\phi$ of weight $w(\phi) =-1$ in the theory. For the gravitational Lagrangian Dirac adopted the expression $L = -\phi ^2R + k D_\mu \phi D^\mu \phi + c\phi ^4 +W_{\mu \nu}W^{\mu \nu}/4$, where $R$
is the Ricci scalar, and $c$ and $k = 6$ are constants. $W_{\mu \nu}$ is the electromagnetic type tensor constructed from the Weyl vector. The Dirac Lagrangian is conformally invariant by construction.

The $f(Q)$ type modified gravity theories are based on the mathematical formalism of the Weyl geometry, by defining the gravitational action $S = \int f(Q) \sqrt{-g} d^4x$ as an arbitrary function of the non-metricity scalar $Q$ \cite{fq1,fq2,fq3,fq4,fq5,fq6,fq7}. For a review of $f(Q)$ gravity see \cite{fq8}.  In the $f(Q)$ theory, and in its extensions, the geometry is solely defined by the non-metricity $Q_{\lambda\mu\nu}$. The investigations of the various models of the theory suggest  that the observational data supports the model's ability to explain the late-time acceleration of the Universe.

An alternative formulation of  Weyl quadratic gravity was proposed in \cite{Gh1,Gh2,Gh3,Gh4,Gh5, Gh6,Gh7, Gh8}, and involves the linearization of the  Lagrangian density of the Weyl quadratic gravity by introducing an auxiliary scalar field. In \cite{Gh4}, the Standard Model (SM) was reconsidered with the inclusion of a Weyl gauge geometry.
 The approach taken was minimal, without introducing any additional degrees of freedom. The gauged scale symmetry D(1), known as Weyl gauge symmetry, is broken spontaneously through a geometric Stueckelberg mechanism \cite{Gh1,Gh2,Gh3,Gh4}. This results in the breaking of the conformal symmetry of the  Weyl quadratic gravity, leading to the emergence of the Einstein-Proca type action, with the inclusion  of the Weyl gauge field $\omega_{\mu}$
\begin{align}
\mathcal{L}_0 &= \sqrt{\tilde{-g}}\big[ -\frac{1}{2}M_P^2\tilde{R}+\frac{3}{4}M_P^2\alpha^2\gamma^2\tilde{\omega}_{\mu}   \tilde{\omega}^{\mu} \nonumber\\
&\qquad-\frac{1}{4}\langle \phi^2\rangle M_P^2 -\frac{1}{4} \tilde{F}^2_{\mu\nu} - \frac{1}{\eta^2}C^2_{\mu\nu\rho\sigma}  \big],
\end{align}
where $M_P$ denotes the Plank mass, $\alpha$ is the Weyl gauge coupling, $\gamma$ and $\eta$ are coupling constants, $\langle\phi\rangle$ is the vacuum expectation value of the scalar field, and $F_{\mu\nu}$ the field strength of the Weyl vector, respectively and $C_{\mu\nu\rho\sigma}$ the Weyl tensor (the tilde represents the transformed fields).

When the mass of the Weyl field $m_{\omega}$ decreases beyond the threshold proportional to the Plank scale $M_P$, $m_{\omega}^2 = \frac{3}{2}q^2 M_P^2$, with $q$ the Weyl gauge coupling \cite{Gh1,Gh2}, the massive gauge field decouples. As a result, a shift from Weyl geometry to Riemannian geometry occurs, which enables the derivation of the Einstein-Hilbert action (the "low-energy" limit of Weyl quadratic gravity), the positive cosmological constant and the Proca action of the gauge field, leading to a viable  quadratic gravity theory.
Likewise, embedding the SM with Weyl geometry has yielded promising results due to the Stueckelberg-Higgs mechanism, notably in the description of inflation (see also \cite{Gh5,Gh6,Gh7}). In the early Universe, the Higgs boson may have been produced through Weyl vector fusion. In addition, in \cite{Gh8} a metric-like Weyl gauge invariant formalism was provided, which is present at the quantum level, and allows the discussion of the Weyl anomaly from another perspective (the conformal symmetry of classical theory is broken at the quantum scale). The physical implications of Weyl geometric gravity were further investigated in \cite{W1, W2,W3,W4,W5,W6,W7,W8,W9}.

An interesting approach to explain dark matter as a geometric effect was considered in \cite{M1},  where the conformal degree of freedom of gravity was isolated by introducing the physical metric in terms of an auxiliary metric and of the gradient of a scalar field, namely
\be
g_{\mu\nu} =  (\tilde{g}^{\alpha\beta}\partial_{\alpha}\phi\partial_{\beta}\phi) \tilde{g}_{\mu\nu}.
\ee
The equation clearly shows that the conformal degree of freedom behaves dynamically, while the theory upholds the invariance with respect to Weyl transformations. After taking the variation of the action of the model
\be
S = -\frac{1}{2} \int [R(g_{\mu\nu}(\tilde{g}_{\mu\nu} ,\phi))+ \mathcal{L}_m] \sqrt{-g(\tilde{g}_{\mu\nu} ,\phi)}d^4x,
\ee
the constraint of the scalar field is found to be
\begin{equation}
g^{\mu\nu} (\partial_{\mu}\phi)(\partial_{\nu}\phi) =  1,
\end{equation}
where the natural units were considered with $c=1$. However, the foundations for mimetic theories were established through the  papers \cite{mm1, mm2, mm3} a few years before, by creating a theory that effectively  describes DE and DM.

Shortly after the proposal of mimetic gravity, the incorporation of Lagrange multipliers $\lambda^{\mu\nu}$ in the action was found to be a reliable approach to obtain the same results \cite{M2}. There are several approaches to the mimetic field theory, besides the Lagrange multiplier, such as disformal transformations \cite{b1,b2}, and obtaining mimetic gravity from the Brans-Dicke Theory \cite{b3}. For a review of mimetic gravity, and its astrophysical and cosmological implications see \cite{Vag}.

An additional notable feature of the mimetic dark matter  is that the introduction of a non-dynamical scalar field could justify inflation, quintessence and a bouncing Universe \cite{M3}. Moreover, the issue of an eternally self-reproducing universe is avoided by coupling the mimetic field to the inflaton potential, as discussed in \cite{M4} where the action
\begin{align}
S &= \int \bigg[ -\frac{1}{2} R + \lambda (g^{\mu\nu}\partial_{\mu}\phi\partial_{\nu}\phi - 1)+\frac{1}{2}g^{\mu\nu}\partial_{\mu}\varphi\partial_{\nu}\varphi \nonumber\\
&\qquad- C(\kappa) V(\varphi) \bigg] \sqrt{-g}d^4x,
\end{align}
was considered, where $\varphi$ is the inflaton field and $C(\kappa)$ a coupling function, with $\kappa \equiv \square \phi$.

Mimetic dark matter was also studied as a ghost free model in \cite{b4}, implying that it emerges as a conformal extension of the Einstein theory with local Weyl invariance. Furthermore,  \cite{mm6} on the ghost free model extends this property to all orders for the mimetic massive graviton. For other investigations in the field of mimetic gravity see \cite{Mi1,Mi2,Mi3,Mi4,Mi5,Mi6,Mi7,Mi8,Mi9,Mi10}.

The main goal of the present work is to investigate the possibility of the implementation of the idea of the mimetic field theory into the framework of the Weyl geometric gravity. We will begin our study by assuming that the gravitational Lagrangian density contains a quadratic term introduced via the Ricci scalar $\tilde{R}^2$, defined in the Weyl geometry, the field strength $\tilde{F}_{\mu\nu}^2$ of the Weyl vector field, and  an effective matter Lagrangian $\mathscr{L}_m$. In view of the fact that we seek to linearize the quadratic action, an auxiliary scalar field is introduced, which allows to rewrite the action as a linear function in the Ricci scalar. The mimetic field is included via the Lagrange multiplier approach in the constructed action, thus leading to the addition of the scalar degree of freedom in the conformal factor (within the Weyl transformations). The mimetic field is constructed by using the same scalar field that was used for the linearization of the quadratic term.  It should be noted that the presence of the mimetic field leads to a class of modified gravity theories, where an additional but constrained scalar degree of freedom is added. Furthermore, the geometrical components, and the variation of the matter part of the action are invariant under Weyl conformal transformations.

The field equations and the constraint of the mimetic field are obtained  after varying the action with respect to the metric tensor, the trace of the Lagrange multipliers, the scalar field, and the Weyl vector, respectively, thus leading to a closed system of field equations, also containing the energy-momentum tensor obtained from the variation of the effective matter Lagrangian.

In order to analyze some specific applications, we need to specify the form of the effective matter Lagrangian. We adopt for the effective matter term a simple form given by the sum of the Lagrangian of the ordinary matter, and of the squares of the Weyl vector and of the scalar field, respectively. Then, the gravitational field equations of the mimetic Weyl gravity are derived for the given form of the effective matter Lagrangian density.

As an application of the mimetic Weyl gravity theory we investigate the cosmological evolution in the Friedmann-Lemaitre-Robertson-Walker (FLRW metric), by considering a time-dependent scalar field and Weyl vector, respectively. We obtain the generalized Friedmann equations for this model, that generalize the standard Friedmann equations of general relativity through the inclusion of the scalar field, of the Weyl vector, and of the Lagrange multiplier. We reformulate the Friedmann equations in a dimensionless form, and we also introduce the redshift representation. We compare the predictions of the model with a small set of observational data for the Hubble function, as well as with the predictions of the standard $\Lambda$CDM model. The free parameters of the model are determined by using the Likelihood analysis applied to the observational data on the Hubble parameter. Our results show that the mimetic Weyl gravity cosmological model  gives a good description of the observational data, and reproduces well the predictions of the $\Lambda$CDM paradigm.

This paper is organized as follows. In Section~\ref{sect1} we briefly discuss the basics of Weyl geometry, and of the Weyl geometric gravity theory. We also introduce  a few fundamental concepts of the mimetic theory. We extend  the Lagrangian of the Weyl geometric gravity by including the mimetic field, and, after some redefinitions of the physical variables we obtain the canonical form of the action. The matter term is added via an effective Lagrangian, depending on the matter Lagrangian, the Weyl vector, and the auxiliary scalar field. The gravitational field equations are obtained by varying the action with respect to the metric tensor. We investigate the cosmological implications of the Mimetic Weyl geometric gravity in Section~\ref{sect2}. By adopting the FLRW metric, we obtain first the generalized Friedmann equations of the model. In order to compare the theoretical predictions with the observational data we reformulate the equations by introducing the redshift representation. We test the model's validity for the description of the late-time expansion by solving numerically the cosmological equations, and by comparing the model with a set of observational values of the Hubble function, and with the standard $\Lambda$CDM model. Finally, we discuss and conclude our results in Section~\ref{sect3}.

\section{From Weyl geometry to mimetic Weyl gravity}\label{sect1}

In the present Section we will first briefly review the foundations of Weyl geometry, and of the Weyl geometric gravity. We will then introduce the basic ideas of mimetic gravity, and we will proceed to the formulation of the theory of mimetic Weyl geometric gravity. After writing down the gravitational action, we obtain the field equations by varying the action with respect to the metric tensor. A specific form of the effective matter Lagrangian is also adopted, and the corresponding field equations are presented.

\subsection{Weyl geometry and Weyl geometric gravity}

The fundamental theorem of Riemannian geometry states that on a Riemannian (non-Euclidian) manifold there is a unique affine connection on the tangent bundle, which is torsion-free and metric compatible, called the Levi-Civita connection. In this geometry, the covariant derivative - the connection between tangent spaces - of the metric tensor identically vanishes. While there is an angle shift of the vectors under parallel transport with respect to the connection, the length of a vector is preserved.  After the derivation of the field equations of General Relativity by Einstein and Hilbert,  in \cite{Weyl1} and \cite{Weyl2} Weyl proposed a long lasting contribution to mathematics and gravitational theories, based on two basic concepts, the conformal gauge symmetry, and the nonmetricity, respectively.

\paragraph{Weyl geometry.} As a generalization of the Riemannian geometry, Weyl geometry is defined by classes of equivalence $(g_{\mu \nu}, \omega_{\alpha})$ of the metric $g_{\mu \nu}$, and of the Weyl vector gauge field $\omega_{\alpha}$, related by the Weyl gauge transformations \cite{Gh3,Gh4,Gh5}
\begin{align}\label{e:transf}
\tilde{g}_{\mu \nu} &= \Sigma^n (x)g_{\mu \nu},\nn\\ \tilde{\omega}_{\mu}& = \omega_{\mu} - \frac{1}{\alpha}\partial_{\mu} \ln \Sigma,\nn\\ \tilde{\phi} &= \Sigma^{-\frac{n}{2}} \phi
\end{align}
where $\Sigma(x)>0$ is the position dependent conformal factor,  $n=1$ is the Weyl charge, $\alpha$ is the Weyl gauge coupling, and $\phi$ is the scalar field.

If a vector of length $l$ is transported by parallel displacement from the point $x^{\mu}$ to $x^{\mu}+\delta x^{\mu}$, in Weyl geometry its change in length is given  by
\begin{equation}
\delta l = l \omega_{\mu}\delta x^{\mu} ,
\end{equation}
 where the new field quantities $\omega_{\mu}$, called the Weyl vector field, satisfy the electromagnetic potential's properties.
 If this vector is transported by parallel displacement round a small closed loop. the total change will be
\begin{equation}
\delta l = l F_{\mu \nu} \delta S^{\mu \nu} ,
\end{equation}
with $\delta S^{\mu \nu}$ denoting the element of area enclosed by the loop and $F_{\mu \nu}$ describing the Weyl field strength of the Weyl vector field.
We can therefore observe that in Weyl geometry there is a formal, geometrical similarity with the electromagnetic fields \cite{Di1,Di2}.

The nonmetricity $Q_{\lambda \mu \nu}$ of the Weyl manifold is obtained as a function of the Weyl vector field, via the covariant derivative of the metric tensor, according to
\begin{equation}
\tilde{\nabla} _{\lambda} g_{\mu\nu} = -\alpha \omega_{\lambda} g_{\mu\nu} = Q_{\lambda \mu \nu}.
\end{equation}

The Weyl connection is a solution to the nonmetricity equation above, where $\tilde{\Gamma}^{\lambda}_{\mu\nu}$ defines $\tilde{\nabla} _{\lambda}$ (we have denoted by a tilde all the quantities defined in the Weyl geometry) with
\begin{equation}
\tilde{\Gamma}^{\lambda}_{\mu\nu} = \Gamma^{\lambda}_{\mu\nu}+\frac{\alpha}{2}\left( \delta^{\lambda}_{\mu} \omega_{\nu} +\delta^{\lambda}_{\nu} \omega_{\mu}-g_{\mu\nu}\omega^{\lambda} \right) ,
\end{equation}
where $\Gamma^{\lambda}_{\mu\nu}$ is the standard Levi-Civita connection associated to the metric $g$. It can be observed that the Weyl field $\omega_{\mu}$ measures the deviation of the Weyl connection (denoted by tilde) from the Levi-Civita connection. It is obvious that if the Weyl vector field tends to zero, $(\omega_{\mu}\rightarrow 0)$, the Weyl geometry becomes the Riemannian geometry. By using $\tilde{\Gamma}^{\lambda}_{\mu\nu}$ one can construct the tensor and scalar curvatures of the Weyl geometry, by using the usual formulae of the Riemannian case, with the curvature tensor $\tilde{R}_{\mu\nu\sigma}^{\lambda}$ being given by
\begin{equation}
\tilde{R}_{\mu\nu\sigma}^{\lambda} =\partial_{\nu} \tilde{\Gamma}^{\lambda}_{\mu\sigma} - \partial_{\sigma} \tilde{\Gamma}^{\lambda}_{\mu\nu} + \tilde{\Gamma}^{\lambda}_{\rho\nu}\tilde{\Gamma}^{\rho}_{\mu\sigma} - \tilde{\Gamma}^{\lambda}_{\rho\sigma}\tilde{\Gamma}^{\rho}_{\mu\nu} .
\end{equation}
The first contraction of the Weyl curvature tensor is
\begin{equation}
\tilde{R}_{\mu\nu} = \tilde{R}_{\mu\lambda\nu}^{\lambda} ,
\end{equation}
and its second contraction, or the Weyl scalar, is given by
\begin{equation}
\tilde{R} = g^{\mu\nu}\tilde{R}_{\mu\sigma} =  g^{\mu\nu} \left( \partial_{\rho} \tilde{\Gamma}^{\rho}_{\mu\nu} - \partial_{\nu} \tilde{\Gamma}^{\rho}_{\mu\rho} + \tilde{\Gamma}^{\rho}_{\mu\nu}\tilde{\Gamma}^{\sigma}_{\rho\sigma} - \tilde{\Gamma}^{\sigma}_{\mu\rho}\tilde{\Gamma}^{\rho}_{\nu\sigma}\right) ,
\end{equation}
and in final form as
\begin{equation}\label{e:Rtilde}
\tilde{R} = R - 3\alpha\nabla_{\mu}\omega^{\mu}-\frac{3}{2}\alpha^2 \omega_{\mu}\omega^{\mu},
\end{equation}
where $R$ is the Ricci scalar and $\nabla_{\mu}\omega^{\lambda} = \partial_{\mu}\omega^{\lambda} + \Gamma^{\lambda}_{\mu\rho}\omega^{\rho}$ is defined in the Riemannian geometry.

Another important geometrical quantity that we can construct based on the Weyl vector is the Weyl field strength $F_{\mu\nu}$ of $\omega_{\mu}$, defined as,% where we assume that the physical property of $F_{\mu\nu}$ is represented as geometry acting on matter \cite{W3},
\begin{equation}\label{e:fieldT}
\tilde{F}_{\mu\nu} = \tilde{\nabla}_{[{\mu}}\omega_{{\nu}]} = \nabla_{[{\mu}}\omega_{{\nu}]}= \nabla_{\mu}\omega_{\nu}-\nabla_{\nu}\omega_{\mu} =  \partial_{\mu}\omega_{\nu} - \partial_{\nu}\omega_{\mu},
\end{equation}
since the Weyl connection $\tilde{\Gamma}^{\lambda}_{\mu\nu} = \tilde{\Gamma}^{\lambda}_{\nu\mu} $ is symmetric.

\paragraph{Weyl geometric gravity.} The gravitational action in Weyl geometry was initially proposed by Weyl in \cite{Weyl1,Weyl2,Weyl3}, and recently reconsidered in \cite{Gh1,Gh2,Gh3} as the simplest conformally invariant Lagrangian density. In the following we also consider in our model an effective matter Lagrangian density $\mathscr{L}_m$ \cite{W5}
\begin{equation}
\mathscr{L}_m = \mathscr{L}_m(L_m,\omega^2,\phi) ,
\end{equation}
where $\mathscr{L}_m$ can depend in general on the ordinary matter Lagrangian $L_m$, on the Weyl vector through $\omega^2=\omega^{\mu}\omega_{\mu}$, and on the scalar fields $\phi$. Therefore, the Lagrangian density of the Weyl geometric gravity theory is given by
\begin{equation}
L_0 = \left[\frac{1}{4!\xi^2}\tilde{R}^2 - \frac{1}{4}\tilde{F}^2_{\mu\nu} + \beta \mathscr{L}_m \right] \sqrt{-\tilde{g}},
\end{equation}
where $\xi<1$ is the perturbative coupling, and $\beta$ is a constant.

We introduce now an auxiliary scalar field $\phi_0$ by extracting from the $\tilde{R}^2$ term (which holds higher derivatives) a scalar degree of freedom, according to the approach pioneered in \cite{Gh1,Gh2,Gh3}. To simplify the calculations, and to linearize the gravitational Lagrangian, we replace $\tilde{R}^2\rightarrow-2\phi_0^2\tilde{R} - \phi_0^4$. The Lagrangian obtained in this way is equivalent to the initial one, since the solution of the equation of motion of the scalar field is $\phi_0^2 = - \tilde{R}$, and after substituting it in the linearized Lagrangian we obtain the initial one. Hence, the linearized Weyl geometric Lagrangian is obtained in the form
\begin{equation}\label{e:L0}
L_0 = \left[-\frac{\phi_0^2}{12\xi^2}\tilde{R} - \frac{\phi_0^4}{4!\xi^2} - \frac{1}{4}\tilde{F}^2_{\mu\nu} + \beta \mathscr{L}_m \right] \sqrt{-\tilde{g}}.
\end{equation}
$L_0$ has a spontaneous breaking to an Einstein-Proca type Lagrangian of the Weyl gauge field \cite{Gh2,Gh3,Gh4}.

By substituting in Eq.~(\ref{e:L0}) the $\tilde{R}$ term from Eq.~(\ref{e:Rtilde}), and by performing a gauge transformation that allows the redefinition of the variables, we obtain a dimensionless action. Therefore, to obtain the action as defined in the Riemannian space, we multiply it with a constant having the dimensions of an action, $1/2\kappa^2_0$, thus obtaining
\begin{align}
S_W &=\frac{1}{2\kappa_0^2} \int\,\bigg[ -\frac{\phi_0^2}{12\xi^2} \left(R-3\alpha \nabla_{\mu} \omega^{\mu}-\frac{3}{2}\alpha^2 \omega_{\mu} \omega^{\mu} \right) \nonumber \\
&\qquad - \frac{\phi_0^4}{4! \xi^2}-\frac{1}{4}\tilde{F}_{\mu \nu}\tilde{F}^{\mu \nu}+ \beta \mathscr{L}_m \bigg] \sqrt{-g}d^4x.
\end{align}

This is the equivalent form of the Weyl quadratic action, linearized in the Ricci scalar.

\subsection{Mimetic gravity}

With the emergence of the "dark universe" concept, the peculiar nature of dark matter suggested that our understanding of the Universe is incomplete, and further motivations were brought to light to modify GR.
In \cite{M1} a reformulation of Einstein's theory of gravity was proposed, where the conformal degree of freedom of gravity was extracted by introducing the physical metric $g_{\mu\nu}$ as given in terms of an auxiliary metric $\tilde{g}_{\mu\nu}$ and a scalar field $\phi_0$, so that
\begin{equation}
g_{\mu\nu} =  (\tilde{g}^{\alpha\beta}\partial_{\alpha}\phi_0\partial_{\beta}\phi_0) \tilde{g}_{\mu\nu},
\end{equation}
where $\tilde{g}^{\alpha\beta}\partial_{\alpha}\phi_0\partial_{\beta}\phi_0$ represents the conformal factor. It is evidently apparent that the physical metric is invariant under conformal transformations of the auxiliary metric $\tilde{g}_{\mu\nu}\rightarrow \Omega^2 \tilde{g}_{\mu\nu}$, $\Omega$ being a function of space-time coordinates. It was shown in \cite{M1} that the extra degree of freedom of the gravitational field could mimic the behavior of cold dark matter at a cosmological scale.

In general terms, the class of modified theories of gravity where an additional scalar degree of freedom is added contain the concept of mimetic gravity. Mimetic DM consists rather of a constrained scalar degree of freedom, than a proper one. Hence, in mimetic gravity the scalar field satisfies the following constraint
\begin{equation}\label{e:gm}
g^{\mu\nu} (\partial_{\mu}\phi_0)(\partial_{\nu}\phi_0) =  \varepsilon.
\end{equation}
Here the constant $\varepsilon$ was introduced to ensure the validity of the units of measurement, since we do not use the natural units as in \cite{M1}.

An equivalent formulation of the mimetic DM model was introduced by \cite{M2}, via the help of Lagrange multipliers $\lambda^{\mu\nu}$, with the obtained equations of motion equivalent to the ones found in \cite{M1}. Therefore, the action of the mimetic gravity model can be written as
\begin{equation}
S_{mimetic}  =\int \lambda(g^{\mu\nu}\partial_{\mu}\phi_0\partial_{\nu}\phi_0-\varepsilon)\sqrt{-g} d^4x,
\end{equation}
where we can replace $\lambda_{\mu\nu}$ since it is fully determined by its trace $\lambda_{\mu\nu}=\lambda(\partial_{\mu}\phi_0)(\partial_{\nu}\phi_0)$.

\subsection{Mimetic Weyl geometric gravity: action and field equations}

We introduce now the total action of the mimetic Weyl geometric gravity, with the mimetic condition added by using the Lagrange multiplier approach, and which is given by
\begin{align}\label{e:action}
\mathcal{S}& = \frac{1}{2\kappa_0^2} \int\,\bigg[ -\frac{\phi_0^2}{12\xi^2} \left(R-3\alpha \nabla_{\mu} \omega^{\mu}-\frac{3}{2}\alpha^2 \omega_{\mu} \omega^{\mu} \right) \nonumber \\
&\qquad - \frac{\phi_0^4}{4! \xi^2}-\frac{1}{4}\tilde{F}_{\mu \nu}\tilde{F}^{\mu \nu}+\lambda(g^{\mu\nu}\partial_{\mu}\phi_0\partial_{\nu}\phi_0-\varepsilon)\nonumber\\
&\qquad+ \beta \mathscr{L}_m \bigg] \sqrt{-g}d^4x.
\end{align}

To bring the Lagrangian to a canonical form, we consider some redefinitions of the physical quantities according to $\phi_0 = \phi\langle\phi_0\rangle\xi$,
$\langle\phi_0\rangle^2/12\beta = 1/2\kappa^2$, where $\langle\phi_0\rangle$ is the vacuum expectation value of $\phi_0$, and $\kappa^2 = 8\pi G/c^4$. Hence we obtain for the action the final form
\begin{align}\label{e:action1}
\mathcal{S}& = \frac{\beta}{2\kappa_0^2} \int\,\bigg[ -\frac{\phi^2}{2\kappa^2} \left(R-3\alpha \nabla_{\mu} \omega^{\mu}-\frac{3}{2}\alpha^2 \omega_{\mu} \omega^{\mu} \right) \nonumber \\
&\qquad - \frac{\xi^2\Lambda}{4\kappa^2}\phi^4 - \frac{1}{2\kappa^2}\frac{3}{\Lambda}F_{\mu \nu}F^{\mu \nu}+ \mathscr{L}_m\nonumber\\
&\qquad + \frac{12}{2\kappa^2}\xi^2\lambda \bigg(g^{\mu\nu}\partial_{\mu}\phi\partial_{\nu}\phi-\frac{\varepsilon}{\xi^2\Lambda}\bigg)\bigg] \sqrt{-g}d^4x,
\end{align}
where we denoted $\Lambda = \langle\phi_0\rangle^2 $.

It should be mentioned that the geometrical part of the action above is invariant under the transformations from Eq.~(\ref{e:transf})
%\begin{equation}\label{e:tr2}
%\tilde{g}_{\mu \nu} = \Sigma g_{\mu \nu}, \tilde{\omega}_{\mu} = \omega_{\mu} - \frac{1}{\alpha}\partial_{\mu} \ln \Sigma,
%\end{equation}
but the matter part should not be necessarily gauge invariant, while its variation must be \cite{transf1,transf2}. Hence, for the variation of the matter action we can write
\begin{align}
\delta S_m = \frac{1}{2}&\int \tilde{T}_{\mu\nu} \delta g^{\mu\nu}\sqrt{-g}d^4x + \int G^{\mu}\delta\omega_{\mu} \sqrt{-g}d^4x\nonumber\\
&+\int\mathcal{F} \delta\phi\sqrt{-g}d^4x,
\end{align}
where $\tilde{T}_{\mu\nu}$ is the effective total energy-momentum tensor, defined as
\begin{equation}\label{e:effEMT}
\tilde{T}_{\mu \nu} = \frac{2}{\sqrt{-g}} \frac{\delta\big(\sqrt{-g}\mathscr{L}_m(L_m,\omega^2,\phi)\big)}{\delta g^{\mu \nu}},
\end{equation}
$G_{\mu}$ is the Weyl current, obtained according to the prescription
\begin{equation}\label{e:current}
G_{\mu} = \frac{\delta \mathscr{L}_m(L_m,\omega^2,\phi)}{\delta\omega^{\mu}},
\end{equation}
and $\mathcal{F}$ is given by
\begin{equation}\label{e:current}
\mathcal{F} = \frac{\delta\mathscr{L}_m(L_m,\omega^2,\phi)}{\delta\phi}.
\end{equation}

Then the variation of Eq.~(\ref{e:action1}) with respect to the metric gives the field equations of the mimetic Weyl geometric gravity theory as
%\begin{widetext}
\bea\label{e:graveq}
&&\Phi \bigg( R_{\mu \nu} - \frac{1}{2} g_{\mu \nu} R\bigg) +\left(g_{\mu \nu}\square - \nabla_{\mu} \nabla_{\nu} \right) \Phi \nonumber\\
&&+\frac{3\alpha}{2} \left( \omega_{\mu}\nabla_{\nu}\Phi+ \omega_{\nu}\nabla_{\mu}\Phi-g_{\mu \nu}\omega^{\rho}\nabla_{\rho} \Phi  \right) \nonumber\\
&& -\frac{3\alpha^2}{2}\Phi  \left( \omega_{\mu}\omega_{\nu}-\frac{1}{2}  g_{\mu \nu}\omega_{\rho} \omega^{\rho}   \right)
 -\frac{1}{4} \xi^2\Lambda \Phi^2 g_{\mu \nu}\nonumber\\
&& - 3\xi^2\lambda \frac{1}{\Phi}(\partial_{\mu}\Phi) (\partial_{\nu}\Phi)
 + \frac{6}{\Lambda} \left( F_{\mu \beta}F_{\nu \alpha} g^{\beta \alpha} -\frac{1}{4} F_{\alpha \beta}^2   g_{\mu \nu}   \right) \nonumber\\
 && = \kappa^2 \tilde{T}_{\mu \nu},
\eea
%\end{widetext}
where we have denoted $\phi^2=\Phi$, and we have also imposed the mimetic condition.

In the previous variation we used the Voss-Weyl formula for the divergence or covariant derivative of a vector field $$\nabla_{\mu}\omega^{\mu} = \frac{1}{\sqrt{-g}} \frac{\partial}{\partial x^{\mu}}\left( \sqrt{-g} \omega^{\mu} \right). $$ We also used the formula of the effective energy-momentum tensor given in Eq.(\ref{e:effEMT}) and the identity $$g^{\mu \nu }\delta R_{\mu \nu }=\left( g_{\mu \nu }\square -\nabla _{\mu
}\nabla _{\nu }\right) \delta g^{\mu \nu }.$$

The constraint of the mimetic field is obtained in the $\Phi$ variable through the variation of the action with respect to $\lambda$, giving
\begin{equation}\label{e:mimetic}
g^{\mu\nu} (\partial_{\mu}\Phi)(\partial_{\nu}\Phi) =  \frac{4\varepsilon}{\Lambda \xi^2}\Phi.
\end{equation}

The contraction of the field equations gives the trace equation
\begin{align}\label{e:l0}
-\Phi R+&3\square \Phi - 3\alpha \omega^{\rho}\nabla_{\rho}\Phi+\frac{3\alpha^2}{2}\omega_{\mu}\omega^{\mu} \Phi-\xi^2\Lambda\Phi^2\nonumber\\
& -3\xi^2\lambda \frac{1}{\Phi}\partial^{\mu}\Phi\partial_{\mu} \Phi=\kappa^2 \tilde{T}.
\end{align}

After the variation of the action Eq.(\ref{e:action1}) with respect to the scalar field $\phi$ we obtain
\begin{align}
\Phi &\left(R-3\alpha \nabla_{\mu} \omega^{\mu}-\frac{3}{2}\alpha^2 \omega_{\mu} \omega^{\mu} \right) + \xi^2\Lambda\Phi^2- \kappa^2\Phi^{1/2} \mathcal{F} \nonumber\\
&\qquad +6\xi^2\lambda\bigg(\square \Phi -\frac{1}{2} \frac{1}{\Phi} \partial_{\mu}\Phi\partial^{\mu} \Phi\bigg)+6\xi^2\nabla_\mu\Phi\nabla^\mu\lambda =0.
\end{align}

Varying the action with respect to the Weyl vector field with the help of the definition of the Weyl field strength from Eq.~(\ref{e:fieldT}), we obtain the equation of motion of $\omega_{\mu}$, similar to a generalized system of Maxwell-Proca equations
\begin{equation}\label{e:gy}
-\frac{4}{\Lambda}\nabla_{\nu} F^{\nu}_{\mu} - \alpha\nabla_{\mu}\Phi+\alpha^2\Phi\omega_{\mu} + \frac{2\kappa^2}{3} G_{\mu} = 0.
\end{equation}

\paragraph{Field equations for $\mathscr{L}_m = L_m + \frac{\gamma}{2} g^{\mu\nu}\omega_{\mu}\omega_{\nu}-\sigma\phi^2$.} For simplicity, in the following we will study the mimetic Weyl geometric gravitational theory by adopting for the  effective matter lagrangian density the form \cite{W5}
\begin{equation}
\mathscr{L}_m\left(L_m, \omega ^2, \phi\right) = L_m + \frac{\gamma}{2} g^{\mu\nu}\omega_{\mu}\omega_{\nu}-\sigma\phi^2.
\end{equation}
Therefore, we can determine the Weyl current $G_{\mu} = \gamma\omega_{\mu}$, $\mathcal{F} = -2\sigma \Phi^{1/2}$, and the effective energy-momentum tensor from Eq.~(\ref{e:effEMT}) as
\begin{equation}\label{e:tilde}
\tilde{T}_{\mu\nu} = T^{(m)}_{\mu \nu}+ \gamma\bigg( \omega_{\mu}\omega_{\nu} -\frac{1}{2} g_{\mu\nu} \omega^2 \bigg)  +\sigma \Phi g_{\mu\nu},
\end{equation}
where $T^{(m)}_{\mu \nu}$ represents the energy-momentum tensor of the ordinary matter
\begin{equation}
T^{(m)}_{\mu \nu} =  \frac{2}{\sqrt{-g}} \frac{\delta\big(\sqrt{-g}L_m\big)}{\delta g^{\mu \nu}}.
\end{equation}

Therefore, with the use of Eq.~(\ref{e:tilde}), the gravitational field equations obtained from the action  Eq.(\ref{e:action1}) become
\begin{widetext}
\begin{align}\label{e:graveq}
\Phi &\bigg( R_{\mu \nu} - \frac{1}{2}g_{\mu \nu} R\bigg) +\left(g_{\mu \nu}\square - \nabla_{\mu} \nabla_{\nu} \right) \Phi+\frac{3\alpha}{2} \left( \omega_{\mu}\nabla_{\nu}\Phi+ \omega_{\nu}\nabla_{\mu}\Phi-g_{\mu \nu}\omega^{\rho}\nabla_{\rho} \Phi  \right)  -\frac{3\alpha^2}{2}\Phi  \left( \omega_{\mu}\omega_{\nu}-\frac{1}{2}  g_{\mu \nu}\omega_{\rho} \omega^{\rho}   \right) \nonumber \\
&\quad -\frac{1}{4} \xi^2\Lambda \Phi^2 g_{\mu \nu} + \frac{6}{\Lambda} \left( F_{\mu \beta}F_{\nu \alpha} g^{\beta \alpha} -\frac{1}{4} F_{\alpha \beta}^2   g_{\mu \nu}   \right) - 3\xi^2\lambda \frac{1}{\Phi}(\partial_{\mu}\Phi) (\partial_{\nu}\Phi) = \kappa^2 \bigg[T^{(m)}_{\mu \nu}+ \gamma\bigg( \omega_{\mu}\omega_{\nu} -\frac{1}{2} g_{\mu\nu} \omega^2 \bigg)  +\sigma \Phi g_{\mu\nu} \bigg].
\end{align}
\end{widetext}

\section{Cosmological applications}\label{sect2}

In the following we will investigate cosmological implications of the mimetic Weyl geometric theory in the isotropic, homogeneous, and flat  Friedmann–Lemaître–Robertson–Walker universe, with the metric given by
\begin{equation}
ds^2 = c^2dt^2 - a^2(t)\left(dx^2+dy^2+dz^2\right),
\end{equation}
where $a(t)$ represents the scale factor. In our calculations we will use the Landau \& Lifshitz \cite{LaLi} sign conventions and definitions.  We also introduce the expansion rate, or the Hubble function, defined as $H = \dot{a}/a$, where a dot denotes the derivative with respect to the cosmological time.

The general form of the energy-momentum tensor of the ordinary matter, valid in an arbitrary frame of reference, is given by
\begin{equation}
T^{(m)}_{\mu\nu} = \bigg(\rho+ \frac{p}{c^2}\bigg)u_{\mu}u_{\nu}- p g_{\mu\nu},
\end{equation}
where $u_{\mu}$ is the four-velocity of matter, $p$ is the thermodynamic pressure, and $\rho$ is the energy density of the baryonic matter. Note that the four-velocity magnitude is fixed such that $u_{\mu}u^{\mu} =c^2$.

The Lagrangian of the ordinary matter is adopted as
\begin{equation}
L_m =  \rho c^2.
\end{equation}

The Weyl vector must be time-dependent only, and its spatial components must vanish, since this is the only case in which the isotropy and the homogeneity of the Universe is preserved. Thus for $\omega _\mu$ we have
\begin{equation}
\omega_{\mu} = \left(\omega_0(t), 0, 0, 0 \right) .
\end{equation}
It is clear from the above equation that the field strength tensor of the Weyl vector identically vanishes, $F_{\mu\nu}\equiv 0$. We will also consider the case where the scalar field only depends on time, $\phi=\phi(t)\rightarrow \Phi = \Phi(t)$.

\subsection{The generalized Friedmann equations}

For the equation of the universe's expansion rate,the first Friedmann equation is found from the (00)-component of Eq.(\ref{e:graveq}),
\begin{align}\label{e:d1}
\frac{3 H^2}{c^2}& + \frac{3 H}{c^2}\frac{\dot{\Phi}}{\Phi}+ \frac{3\alpha}{2c}\omega_0\frac{\dot{\Phi}}{\Phi}-\frac{3\alpha^2}{4}\omega_0^2-\frac{1}{4}\xi^2\Lambda \Phi\nonumber\\
&\qquad - \frac{3\xi^2\lambda}{c^2}\frac{\dot{\Phi}^2}{\Phi^2} = \frac{\kappa^2}{\Phi} \bigg(\rho c^2+\frac{1}{2}\gamma\omega_0^2+\sigma\Phi\bigg).
\end{align}

The second Friedmann equation for $\mu=1,2,3$ giving the Universe's acceleration rate is given by
\begin{align}\label{e:d2}
\frac{1}{c^2}&(2\dot{H}+3H^2) +\frac{1}{c^2}\bigg( \frac{\ddot{\Phi}}{\Phi} +2H  \frac{\dot{\Phi}}{\Phi} \bigg) - \frac{3\alpha}{2c}\omega_0 \frac{\dot{\Phi}}{\Phi} \nonumber\\
&\qquad+\frac{3\alpha^2}{4}\omega_0^2-\frac{\xi^2\Lambda}{4}\Phi = -\frac{\kappa^2}{\Phi}\bigg( p+\frac{1}{2}\gamma\omega_0^2-\sigma\Phi \bigg).
\end{align}

The scalar field equation becomes
\begin{align}\label{e:d3}
 &-\frac{6}{c^2}(\dot{H}+2H^2)  - \frac{3\alpha}{c}(\dot{\omega}_0+3H\omega_0)-\frac{3\alpha^2}{2} \omega_0^2+\xi^2\Lambda\Phi  \nonumber\\
&\qquad + 2\kappa^2\sigma+\frac{6\xi^2\lambda}{c^2}\bigg(3H \frac{\dot{\Phi}}{\Phi}-\frac{1}{2}\frac{\dot{\Phi}^2}{\Phi^2}+\frac{\ddot{\Phi}}{\Phi}\bigg)=0 .
\end{align}

From the mimetic field constraint, given by Eq.~(\ref{e:mimetic}), taking for $\Phi$ only the positive values, we find for the scalar field the expression
\begin{equation}
\Phi =\frac{\varepsilon c^2}{\Lambda \xi^2}(t-t_0)^2,
\end{equation}
where $t_0$ represents a constant of integration.

From Eq.~(\ref{e:gy}), we obtain the vector field equation as
\begin{equation}
-\frac{\alpha}{c}\dot{\Phi} +\alpha^2 \Phi \omega_0 + \frac{2\kappa^2\gamma}{3}\omega_0=0,
\end{equation}
from which we find the form of the Weyl vector in terms of the scalar field  $\Phi$ and its derivative as
\begin{equation}
\omega_0 = \frac{\alpha\dot{\Phi}}{c\bigg(\alpha^2\Phi+\frac{2\kappa^2\gamma}{3}\bigg)}.
\end{equation}

Hence, the system of the generalized Friedmann equations given by  Eqs.(\ref{e:d1}) and (\ref{e:d2}) becomes %, (\ref{e:d3})

\begin{align}\label{e:d11}
&\frac{3 H^2}{c^2} + \frac{3 H}{c^2}\frac{\dot{\Phi}}{\Phi}+\frac{3\alpha^2}{4c^2}\frac{\dot{\Phi}^2}{\Phi^2}\bigg( \alpha^2 + \frac{2\kappa^2\gamma}{3\Phi}\bigg)^{-1}\nonumber\\
&\qquad -\frac{\xi^2\Lambda}{4} \Phi- \frac{3\xi^2\lambda}{c^2}\frac{\dot{\Phi}^2}{\Phi^2} = \frac{\kappa^2}{\Phi} \rho c^2+\kappa^2\sigma,
\end{align}

\begin{align}\label{e:d22}
&\frac{1}{c^2}(2\dot{H}+3H^2) +\frac{1}{c^2}\bigg( \frac{\ddot{\Phi}}{\Phi} +2H  \frac{\dot{\Phi}}{\Phi} \bigg) -\frac{1}{4}\xi^2 \Lambda \Phi \nonumber\\
&\qquad- \frac{3\alpha^2}{4c^2}\frac{\dot{\Phi}^2}{\Phi^2}\bigg( \alpha^2 + \frac{2\kappa^2\gamma}{3\Phi}\bigg)^{-1}=-\frac{\kappa^2 p}{\Phi}+\kappa^2\sigma.
\end{align}

\begin{comment}
\begin{align}\label{e:d33}
 \dot{H}& +2H^2+\bigg[3H (t-t_0) +\alpha \bigg] \bigg[(t-t_0)^2 +\frac{2\kappa^2\gamma}{3\alpha^2}\frac{\Lambda \xi^2}{\varepsilon c^2}\bigg]^{-1}+ \nonumber\\
& +(t-t_0)^2(1-2\alpha)\bigg[(t-t_0)^2 +\frac{2\kappa^2\gamma}{3\alpha^2}\frac{\Lambda \xi^2}{\varepsilon c^2}\bigg]^{-2}- \nonumber\\
&  -\frac{\varepsilon c^4}{6}  (t-t_0)^2 -\frac{2\kappa^2\sigma}{3c}\bigg(\frac{\Lambda\xi^2}{\varepsilon}\bigg)^{3/2} (t-t_0)^{-3}-\frac{6H\xi^2\lambda}{t-t_0}=0,
\end{align}
\end{comment}

\begin{comment}
The density parameters are defined as follows, the ordinary matter is attributed to radiation and ?baryons or dust particles?, dark matter is attributed to the mimetic field and dark energy to the scalar field,
\begin{equation}
\Omega_m= \frac{\kappa^2 \rho}{3H^2(\pm t + C)^2}
\end{equation}
\begin{equation}
\Omega_ {DM}= -\frac{2\kappa^2 \lambda}{3H^2(\pm t + C)^2}
\end{equation}
\begin{equation}
\Omega_{DE}= \frac{\kappa^2 V(t,H)}{3H^2(\pm t + C)^2}
\end{equation}
\end{comment}

\subsubsection{Dimensionless form}

To minimize the number of free parameters of the model we now introduce the set of dimensionless variables $\left(h,\tau,r,P,\tilde{\Lambda},\varepsilon _0,\gamma _0,\sigma _0\right)$, defined according  to
\bea
H&=&H_0h ,	\quad  \;\;t=\frac{1}{H_0}\tau , 	\quad  \;\;\;\;t_0=\frac{1}{H_0} \tau_0, \nonumber\\
\rho&=&\frac{H_0^2}{8\pi G} r,	\quad 	 p=\frac{H_0^2c^2}{8\pi G } P ,	\quad	\Lambda=\frac{H_0^2}{c^2} \tilde{\Lambda}, \nonumber\\
\varepsilon&=&\frac{H_0^4}{c^4} \varepsilon_0,	\quad \gamma=\frac{c^4}{8\pi G}\gamma_0,	\quad	\sigma=\frac{H_0^2c^2}{8\pi G}\sigma_0.
\eea

The set of the generalized Friedmann equations  Eqs.(\ref{e:d11})-(\ref{e:d22}) takes the dimensionless form
\begin{align}\label{r0}
&3h^2 +3h\frac{1}{\Phi}\frac{d\Phi}{d\tau} +\frac{3\alpha^2}{4}\frac{1}{\Phi^2} \bigg(\frac{d\Phi}{d\tau}\bigg)^2\bigg(\alpha^2+\frac{2\gamma_0}{3\Phi}\bigg)^{-1}  \nonumber\\
&\qquad-\frac{1}{4}\xi^2\tilde{\Lambda} \Phi-3\xi^2\lambda \frac{1}{\Phi^2} \bigg(\frac{d\Phi}{d\tau}\bigg)^2 = \frac{r}{\Phi} +\sigma_0,
\end{align}
\begin{align}\label{e:d111}
&2\frac{dh}{d\tau} + 3h^2 +\frac{1}{\Phi}\frac{d^2\Phi}{d\tau^2}+2h\frac{1}{\Phi}\frac{d\Phi}{d\tau} -\frac{1}{4}\xi^2\tilde{\Lambda} \Phi\nonumber\\
&\qquad -\frac{3\alpha^2}{4}\frac{1}{\Phi^2} \bigg(\frac{d\Phi}{d\tau}\bigg)^2\bigg(\alpha^2+\frac{2\gamma_0}{3\Phi}\bigg)^{-1} = -\frac{P}{\Phi}+\sigma_0.
\end{align}

In the dimensionless variables introduced above, we have
\begin{equation}
\Phi(\tau)=\frac{\varepsilon_0}{\xi^2 \tilde{\Lambda}} \tau^2 ,
\end{equation}
or
\be
 \frac{2\gamma_0}{3\Phi} =  \frac{\delta_0}{\tau^2},
\ee
where we assumed that $\tau_0=0$, and we have denoted $\beta=2\gamma_0\tilde{\Lambda}/3\varepsilon_0\xi^2$. Considering now the pressureless case $P=0$, the dynamical equation, Eq.~(\ref{e:d111}) becomes
\begin{align}\label{e:d1111}
&2\frac{dh}{d\tau} + 3h^2 +\frac{2}{\tau^2}+\frac{4h}{\tau}-\frac{\varepsilon_0}{4} \tau^2 -\frac{3\alpha^2}{\tau^2}\bigg(\alpha^2+ \frac{\beta}{\tau^2}\bigg)^{-1} =\sigma_0.
\end{align}

Eq.~(\ref{e:d1111}) can be reformulated as
\be
2\frac{HD}{d\tau} + 3h^2=-P_{eff},
\ee
where we have introduced the effective pressure of the dark energy $P_{eff}$, defined as
\be
P_{eff}=\frac{2}{\tau^2}+\frac{4h}{\tau}-\frac{\varepsilon_0}{4} \tau^2 -\frac{3\alpha^2}{\tau^2}\bigg(\alpha^2+ \frac{\beta}{\tau^2}\bigg)^{-1}-\sigma_0.
\ee

On the other hand Eq.~(\ref{r0}) can be reformulated as
\be
3h^2=\frac{\xi ^2\tilde{\Lambda}}{\epsilon _0}\frac{r}{\tau ^2}+r_{eff},
\ee
where we have denoted
\be
r_{eff}=\frac{\varepsilon _0}{4}\tau^2+\frac{12\lambda \xi^2}{\tau ^2}-\frac{6h}{\tau}-\frac{3\alpha ^2}{\tau ^2}\left(\alpha ^2+\frac{\beta }{\tau^2}\right)^{-1}+\sigma _0.
\ee

The Weyl vector can be obtained as
\be
\omega _0=\frac{2\alpha H_0}{c}\frac{1}{\tau \left(\alpha ^2+\frac{\beta }{\tau ^2}\right)}.
\ee

We also consider the parameter $w_{eff}$ of the equation of state of the effective, geometric dark energy, considered  as a perfect fluid, defined as
\begin{equation}
w_{eff} = \frac{P_{eff}}{r_{eff}}.
\end{equation}

\subsubsection{Redshift representation}

To facilitate the comparison with the observational data we introduce, instead of the cosmological time,  the redshift variable $z$, defined as
\begin{equation}
1+z=\frac{1}{a}.
\end{equation}
We can now replace in the Friedmann equations the derivative with respect to the time with the derivative with respect to $z$, by using the mathematical identity
\be
\frac{d}{d\tau} = \frac{d}{dz}\frac{dz}{d\tau}=-(1+z)h(z)\frac{d}{dz}.
\ee
Hence, in the redshift space, the generalized Friedmann equations, describing the cosmological dynamics in the mimetic Weyl geometric gravity take the final form
\begin{align}\label{e:f1}
&-2(1+z)h(z)\frac{dh(z)}{dz} + 3h^2(z) +\frac{2}{\tau^2(z)}+\frac{4h(z)}{\tau(z)}\nonumber\\
& -\frac{\varepsilon_0}{4} \tau^2(z)-\frac{3\alpha^2}{\tau^2(z)}\bigg(\alpha^2+\frac{\beta}{\tau^2(z)}\bigg)^{-1}-\sigma_0=0.
\end{align}
\begin{equation}\label{e:67}
\frac{d\tau}{dz} = -\frac{1}{(1+z)h(z)}.
\end{equation}

These equations describe the dynamical evolution of the Hubble function with respect to the redshift. In order to solve the system of equations, they must be integrated numerically with the initial conditions $h(0)=1$, and $\tau(0)=\tilde{\tau} _0$, respectively. Once $h(z)$ and $\tau (z)$ are known, from the first Friedmann equation (\ref{r0}) one obtains the ordinary matter density evolution, $r=r(z)$.

To describe the decelerating/accelerating nature of the cosmic expansion we introduce the deceleration parameter $q$, defined as
\begin{equation}
q = -1 + \frac{d}{d\tau} \frac{1}{h}=-1-(1+z)\frac{1}{h(z)}\frac{dh(z)}{dz}.
\end{equation}

\subsection{Numerical results}

\paragraph{Best fit values of the model parameters.} In order to find the best fit value of the parameters $H_0$, $\bar\tau_0$, $\alpha$, $\beta$, $\epsilon_0$ and $\sigma_0$, we use the Likelihood analysis using the observational data on the Hubble parameter in the redshift range $z\in(0.07,2.36)$  tabulated in \cite{hubble1}.

In the case of independent data points, the likelihood function can be defined as
\begin{align}\label{68}
	L=L_0e^{-\chi^2/2},
\end{align}
where $L_0$ is the normalization constant, and the quantity $\chi^2$ is defined as
\begin{align}\label{69}
	\chi^2=\sum_i\left(\frac{O_i-T_i}{\sigma_i}\right)^2.
\end{align}
Here $i$ counts the data points, $O_i$ are the observational value, $T_i$ are the theoretical values, and $\sigma_i$ are the errors associated with the $i$th data obtained from observations.

By maximizing the likelihood function, the best fit values of the parameters $H_0$, $\sigma_0$, $\bar\tau_0$, $\epsilon_0$, $\alpha$ and $\beta$ at $1\sigma$ confidence level, can be obtained as

\begin{figure*}
	\includegraphics[scale=0.4]{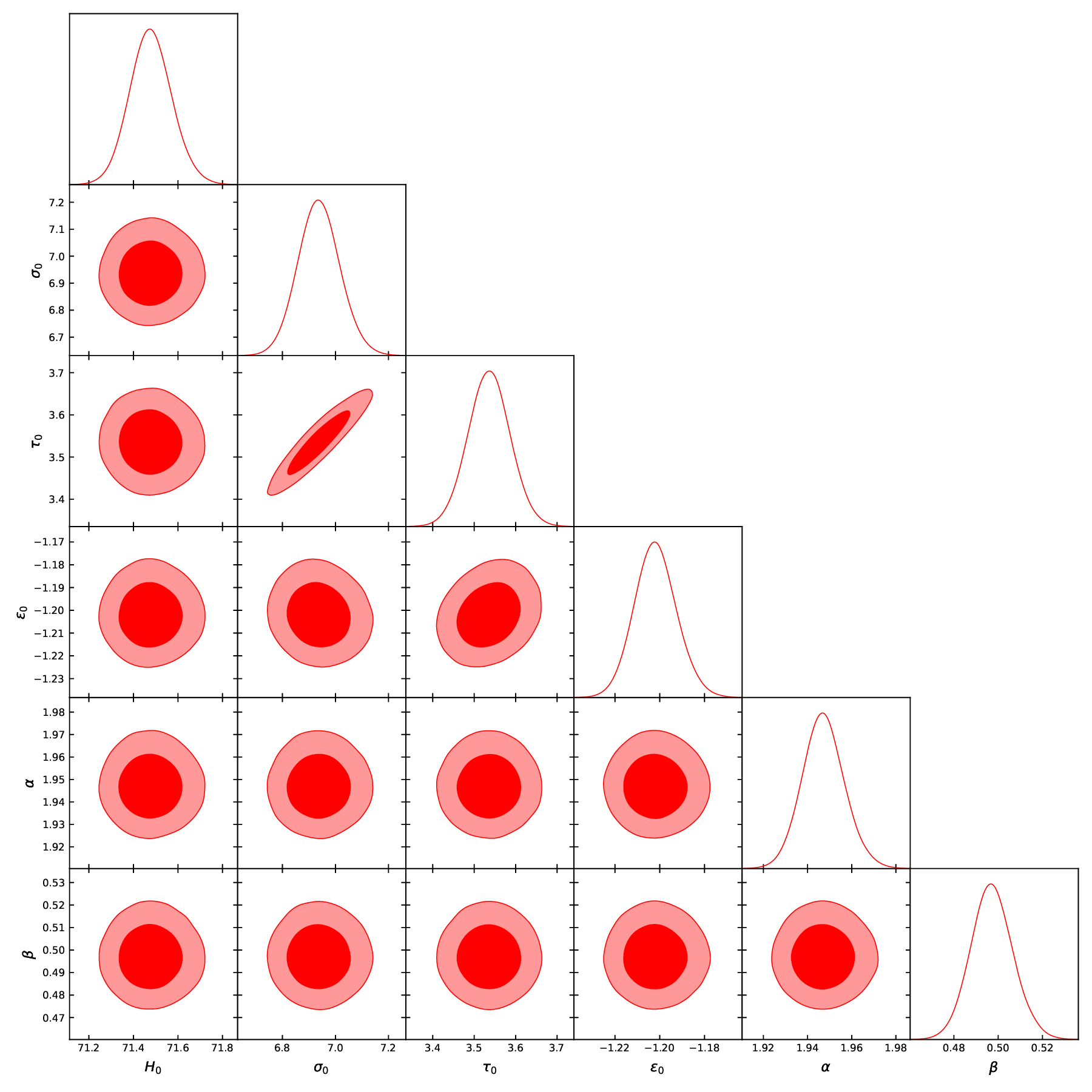}
	\caption{The corner plot for the values of the parameters $H_0$, $\sigma_0$, $\bar\tau_0$, $\epsilon_0$, $\alpha$ and $\beta$ with their $1\sigma$ and $2\sigma$ confidence levels for the Mimetic-Weyl gravity model. \label{cornerplot}}
\end{figure*}

\begin{align}\label{bestfit}
	H_0&=71.476^{+0.095}_{-0.091},\;
	\sigma_0=6.937^{+0.079}_{-0.077},\nonumber\\
	\bar\tau_0&=3.536^{+0.050}_{-0.050}, \;\;\;
	\epsilon_0=-1.202^{+0.010}_{-0.009},\nonumber\\
	\alpha&=1.947^{+0.010}_{-0.009},\;\;\;
	\beta=0.497^{+0.010}_{-0.009}.
\end{align}

The corner plot for the values of the parameters $H_0$, $\sigma_0$, $\bar\tau_0$, $\epsilon_0$, $\alpha$ and $\beta$ with their $1\sigma$ and $2\sigma$ confidence levels is shown in Fig.~\ref{cornerplot}.

The differences between the Mimetic Weyl geometric gravity cosmological model and the $\Lambda CDM$ standard paradigm can also be seen through the values of the chi-squared function for the Hubble function, which are shown in Table~\ref{table1}.

\begin{table}[h!]
\centering
\begin{tabular}{|c|c|}
\hline
Model & $\chi^2$  \\\hline\hline
$\Lambda CDM$ & 21.127 \\\hline
Mimetic-Weyl  & 24.622  \\\hline
\end{tabular}
\caption{The $\chi^2$ for the Hubble function for the Mimetic-Weyl gravity and the $\Lambda CDM$ model.}\label{table1}
\end{table}

\subsubsection{Cosmological parameters}

\begin{figure*}
	\includegraphics[scale=0.57]{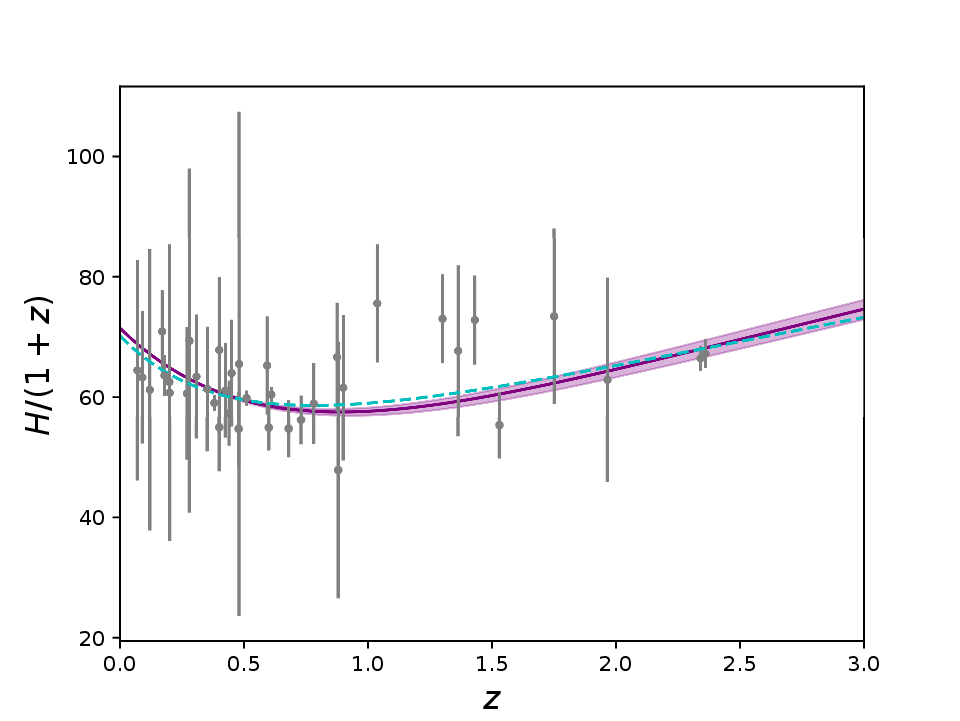}\includegraphics[scale=0.57]{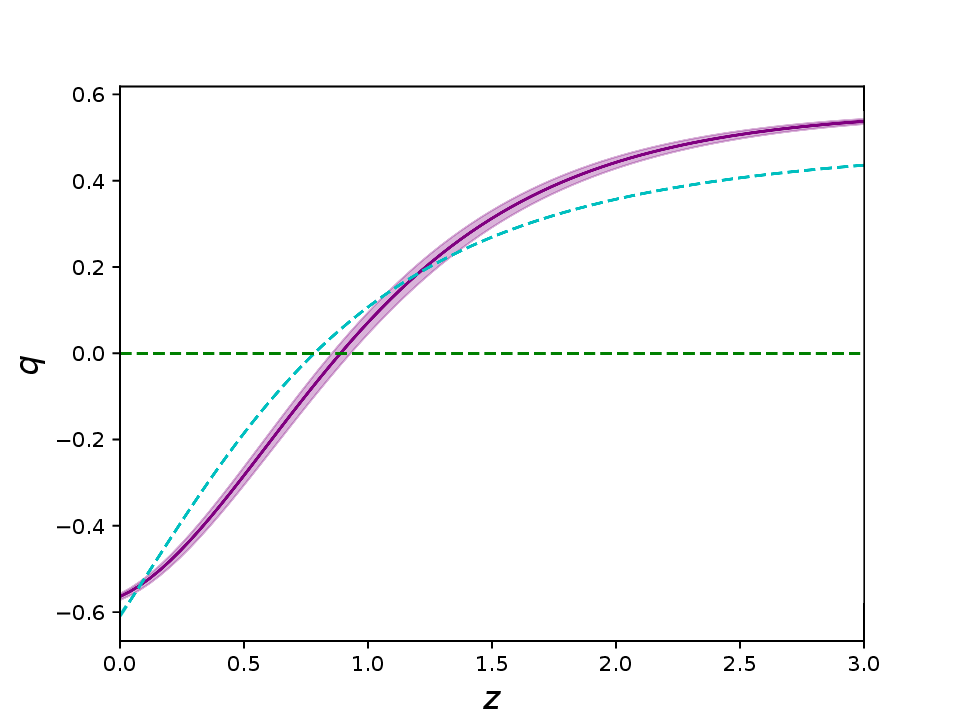}
	\caption{\label{fighubq} The behavior of the resealed Hubble parameter $H(z)/(1+z)$ (left panel) and of the deceleration parameter $q(z)$ (right panel) as a function of the redshift $z$  for the Mimetic Weyl gravity model for the best fit values of the parameters, as given by Eqs.~(\ref{bestfit}). The shaded area denotes the $1\sigma$ error. The dashed line represents the $\Lambda$CDM model.}
\end{figure*}

\paragraph{Hubble function, deceleration parameter, matter density, cosmic time.} The redshift evolution of the Hubble function and of the deceleration parameter $q$ are represented, for the Mimetic Weyl geometric model, in Fig.~\ref{fighubq}. As one can see from left panel of Fig.~\ref{fighubq}, the present cosmological model gives a good description of the observational data for the Hubble function, and reproduces almost perfectly the predictions of the $\Lambda$CDM model up to a redshift $z=3$. The deceleration parameter, presented in the right panel of Fig.~\ref{fighubq} shows some small differences with respect to $q$ as predicted by $\Lambda$CDM. At higher redshanks, the Mimetic Weyl geometric gravity models has slightly higher values of $q$, while at lower redshanks they are slightly smaller than those obtained from $\Lambda$CDM. There is a small difference between models in the numerical value of the transition redshift $z_{tr}$, defined as $q\left(z_{tr}\right)=0$, the transition to an accelerating phase occurring at a slightly higher redshift as in the $\Lambda$CDM model.

\begin{figure*}
	\includegraphics[scale=0.57]{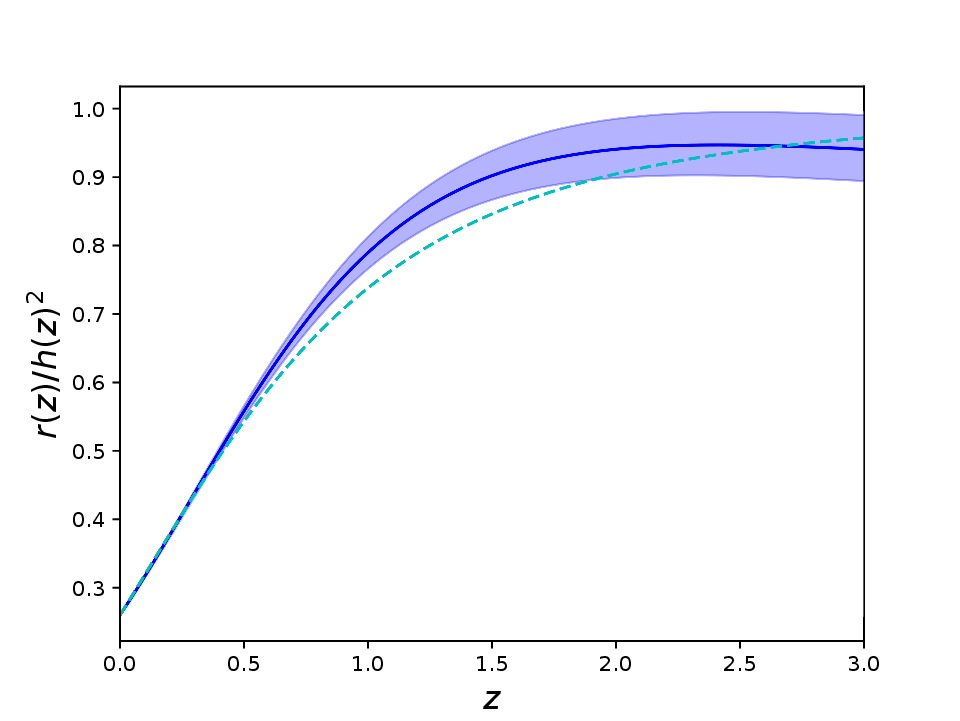}\includegraphics[scale=0.57]{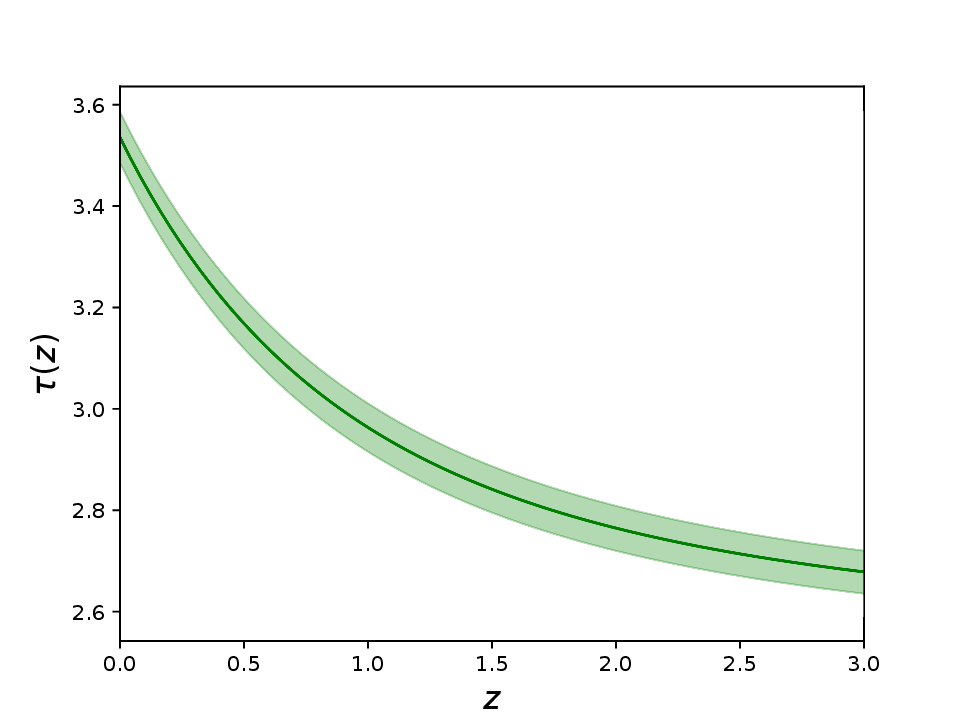}
	\caption{\label{figqj} The behavior of the matter dominance $r(z)/h^2(z)$ (left panel) and of the dimensionless time coordinate $\tau$ (right panel) as a function of the redshift $z$ for the Mimetic-Weyl gravity model for the best fit values of the parameters as given by Eqs.~(\ref{bestfit}). The shaded area denotes the $1\sigma$ error. The dashed line represents the $\Lambda$CDM model.}
\end{figure*}

The behavior of the matter dominance $r(z)/h^2(z)$ term and of the cosmological time are represented in Fig.~\ref{figqj}. There is a good concordance between the matter density predictions, presented in the left panel of Fig.~\ref{figqj}, in Mimetic Weyl gravity, and $\Lambda$CDM. Up to a redshift of around $z=0.5$ the predictions of the two models basically agree, but at redshanks $z>0.5$ the Mimetic Weyl gravity models predicts slightly higher matter densities.  It should be noted that the constant $\gamma_0$ was not fitted in the numerical fitting procedure,  and thus it remains arbitrary. This is why we  consider the resealed $\Phi/\gamma_0$ function. This also happens for the matter dominance $r(z)/h^2(z)$ in Eq.~\eqref{r0}. In order to make the energy density comparable with the $\Lambda CDM$ model, we fix the value of the parameter $\gamma_0$ so that the value of the energy density at the present time $r(z=0)=\Omega_{m0}$ becomes equal to the value of the standard $\Lambda CDM$ value. As a result we obtain $\gamma_0=0.019$, which is used in the Figures representing $\Phi$ and $r/h^2$.

The behavior of the cosmic time as a function of the redshift is depicted in the right panel of Fig.~\ref{figqj}, and it shows a monotonic, and nonlinear dependence of the cosmological time $\tau$ on the redshift.

\paragraph{Statelier diagnostic-jerk and snap parameters.} The Taylor series expansion of the scale factor can be generally represented as  \cite{SJ}
\bea
a(t)&=&a_0\Bigg\{1+\frac{1}{1!}H_0\left(t-t_0\right)-\frac{1}{2!}q_0H_0^2\left(t-t_0\right)^2\nonumber\\
&&+\frac{1}{3!}j_0H_0^3\left(t-t_0\right)^3+\frac{1}{4!}s_0H_0^4\left(t-t_0\right)^4\nonumber\\
&&+O\left[\left(t-t_0\right)^5\right]\Bigg\},
\eea
where the jerk and snap parameters are defined as
\be
j=\frac{1}{H^3}\frac{1}{a}\frac{d^3a}{dt^3},\qquad s=\frac{1}{H^4}\frac{1}{a}\frac{d^4a}{dt^4}.
\ee
In terms of the deceleration parameter $j$ and $s$ can be obtained as \cite{SJ}
\be
j(z)=q(z)+2q^2(z)+(1+z)\frac{dq(z)}{dz},
\ee
and
\be
s(z)=-(1+z)\frac{dj(z)}{dz}-2j(z)-3j(z)q(z),
\ee
respectively.

\begin{figure*}
	\includegraphics[scale=0.57]{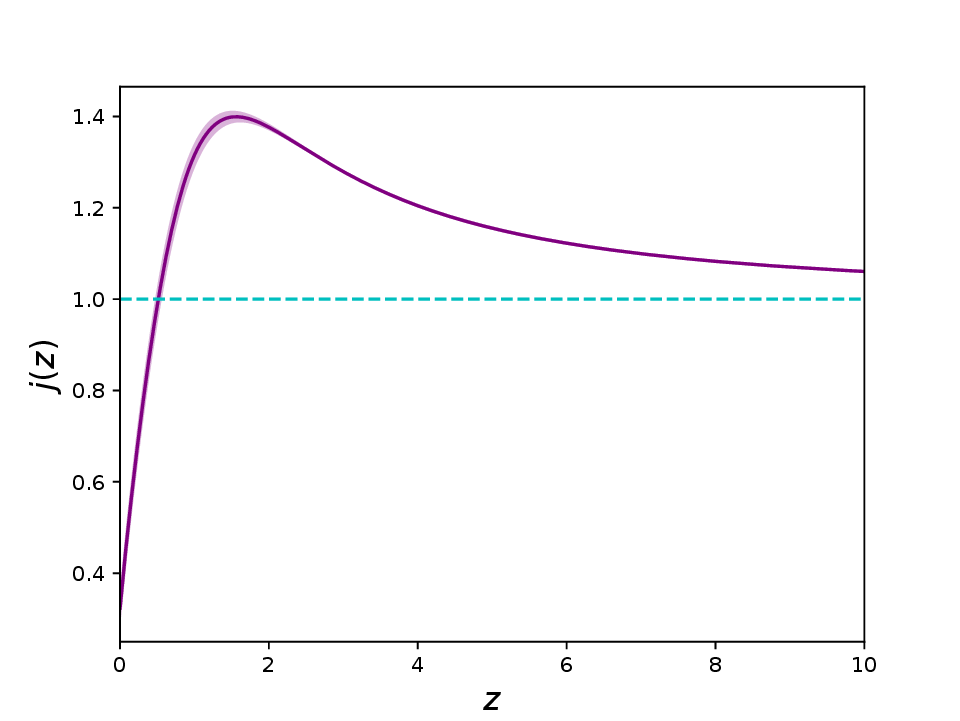}\includegraphics[scale=0.57]{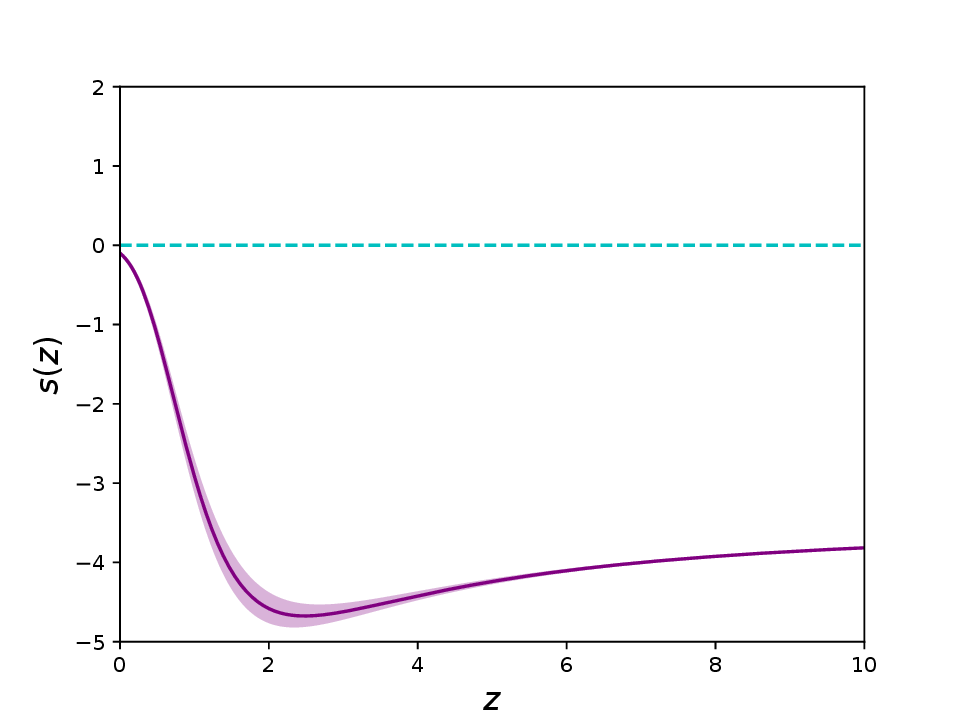}
	\caption{\label{figss} The evolution of the jerk parameter $j(z)$ (left panel) and of the snap parameter $s(z)$ (right panel) as a function of the redshift $z$ in the Mimetic-Weyl gravity model for the best fit values of the parameters as given by Eqs.~(\ref{bestfit}). The shaded area denotes the $1\sigma$ error. The dashed line represents the $\Lambda$CDM model.}
\end{figure*}

The redshift evolution of the jerk and snap parameters are shown, for the Mimetic Weyl geometric gravity model, in Fig.~\ref{figss}. For the $\Lambda$CDM model $j$ and $s$ take the constant values $j=1$, and $s=0$, respectively. As one can see from both panels of Fig.~\ref{figss}, in the Mimetic Weyl gravity theory $j$ and $s$ differ significantly from their $\Lambda$CDM counterparts. $j(z)$, which takes only positive values,  has a maximum value at $z\approx 2$, while $s(z)$, evolving in the negative domain, has a minimum at the same redshift.

\begin{figure*}
	\includegraphics[scale=0.57]{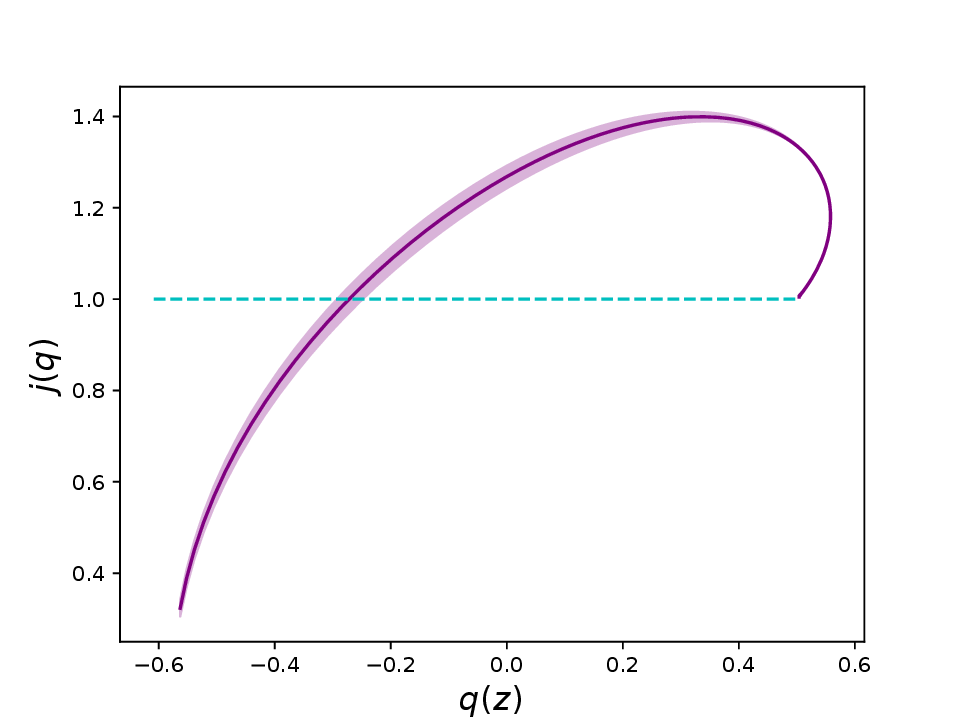}\includegraphics[scale=0.57]{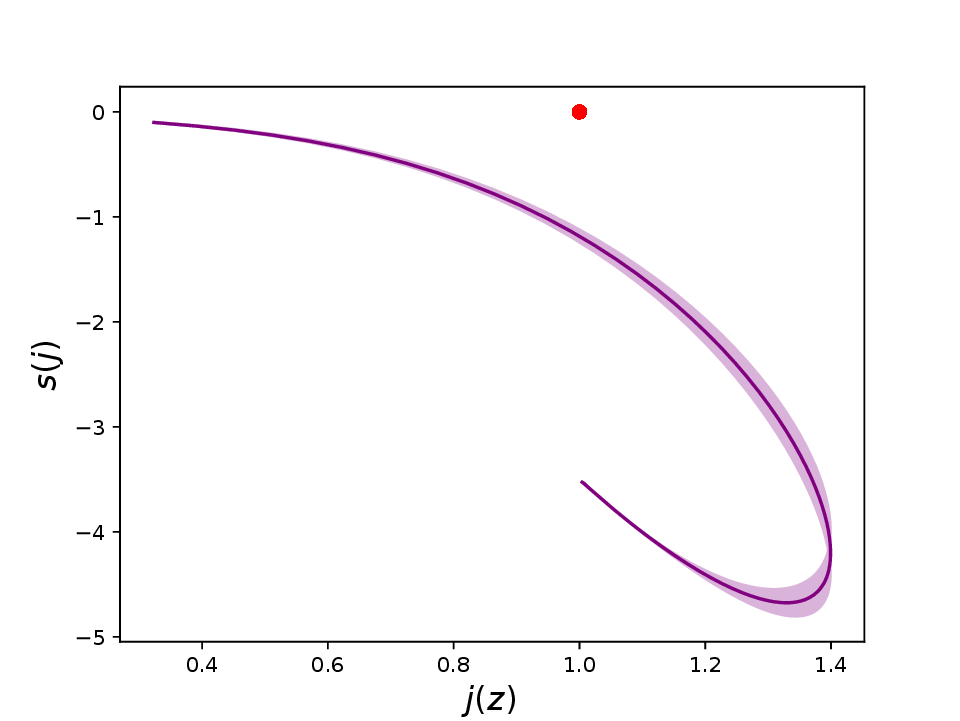}
	\caption{\label{figjqo} The behaviour of the jerk parameter $j$ as a function of the deceleration parameter $q$, $j=j(q)$ (left panel), and of the snap parameter as a function of the jerk parameter, $s=s(j)$ (right panel) for the Mimetic Weyl gravity model for the best fit values of the parameters as given by Eqs.~(\ref{bestfit}). The shaded area denotes the $1\sigma$ error. The dashed line and the red dot represent $\Lambda$CDM model.}
\end{figure*}

The dependence of the jerk parameter $j$ on the deceleration parameter $q$, and the dependence of $s$ on $j$ are presented  for the Mimetic Weyl geometric gravity model, in Fig.~\ref{figjqo}. We find again a significant difference between the behaviors of these parameters in the present model, as compared to the $\Lambda$CDM model.

\paragraph{Geometric quantities - $r_{eff}$, $p_{eff}$, $w_{eff}$, Lagrange multiplier, Weyl vector, scalar field.} 

\begin{figure*}
	\includegraphics[scale=0.57]{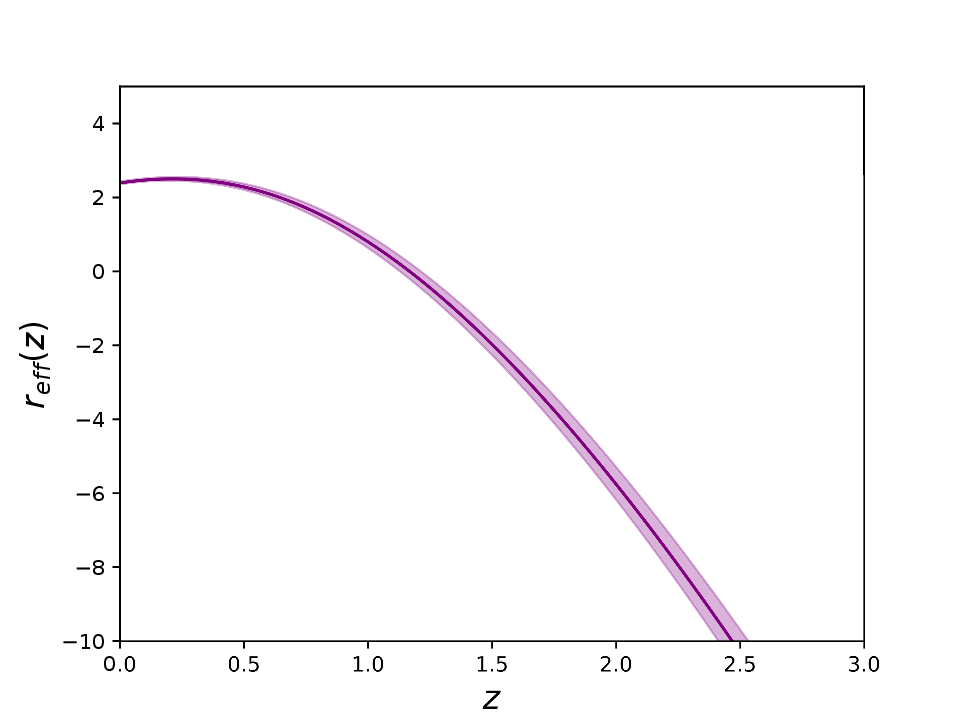}\includegraphics[scale=0.57]{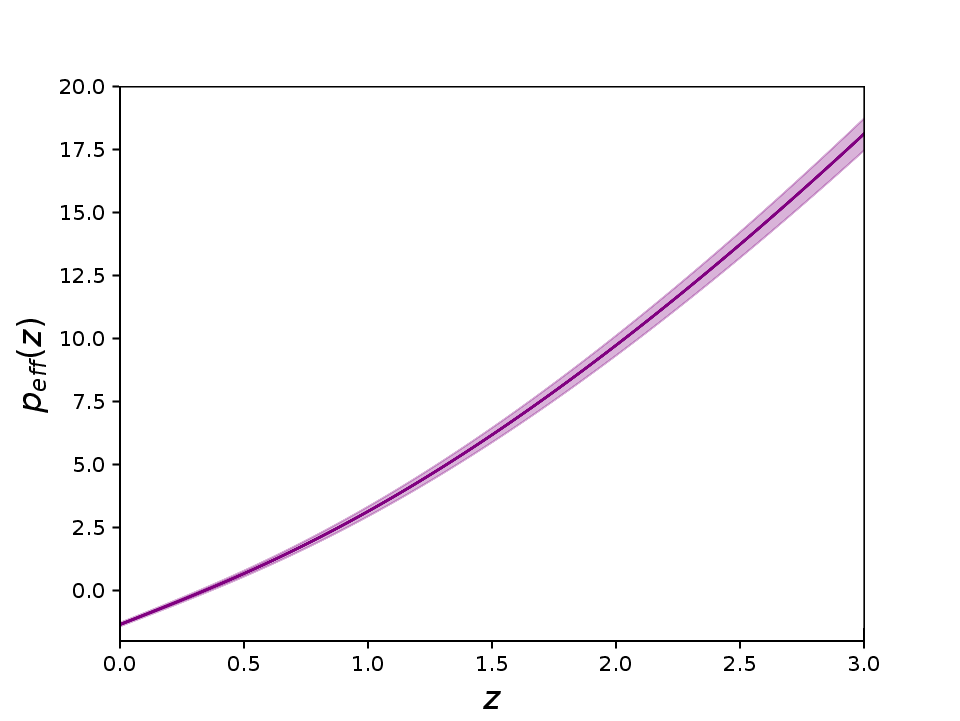}
	\caption{\label{figrpeff} Redshift evolution of the effective geometric energy density $r_{eff}(z)$ (left panel) and of the effective geometric pressure  $P_{eff} (z)$ (right panel) as a function of the redshift for the Mimetic Weyl gravity model for the best fit values of the parameters as given by Eqs.~(\ref{bestfit}). The shaded area denotes the $1\sigma$ error.}
\end{figure*}

The redshift variation of the effective geometric energy density and pressure of the cosmological fluid are represented in Fig.~\ref{figrpeff}. The effective energy density is a decreasing function of the redshift, which takes positive values in the redshift range $z\in (0, 1.25)$, and it becomes zero for $z\approx 1.25$. For $z>1.25$, $r_{eff}$ becomes negative. On the other hand, the effective pressure, a monotonically increasing function of the redshift, is positive for all $z$, except and initial range of $z\in (0.0.25)$. 

\begin{figure*}
	\includegraphics[scale=0.57]{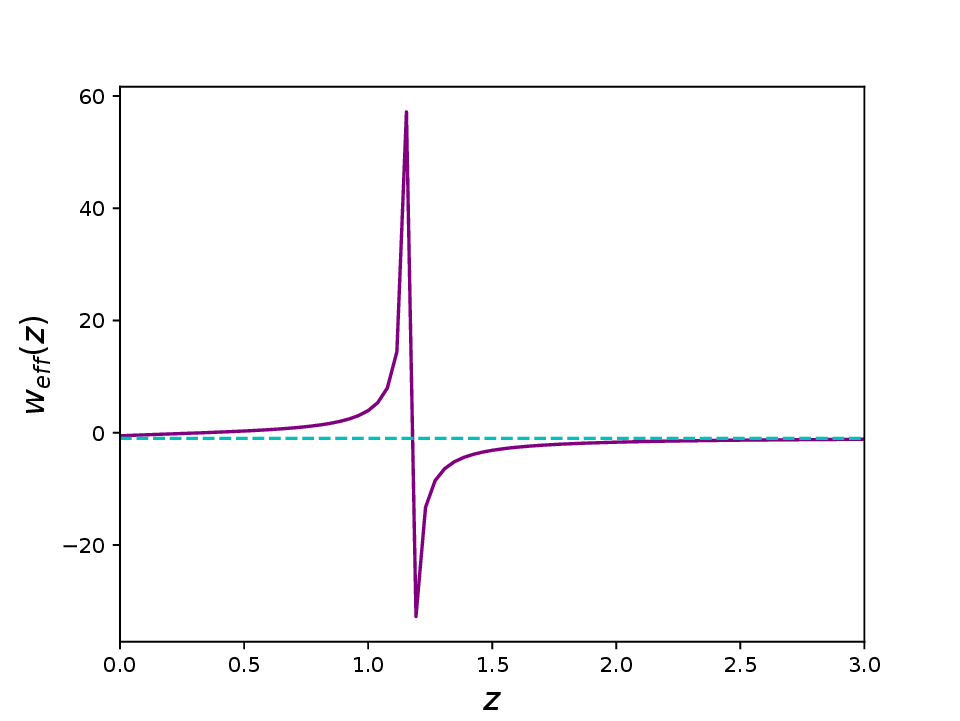}\includegraphics[scale=0.57]{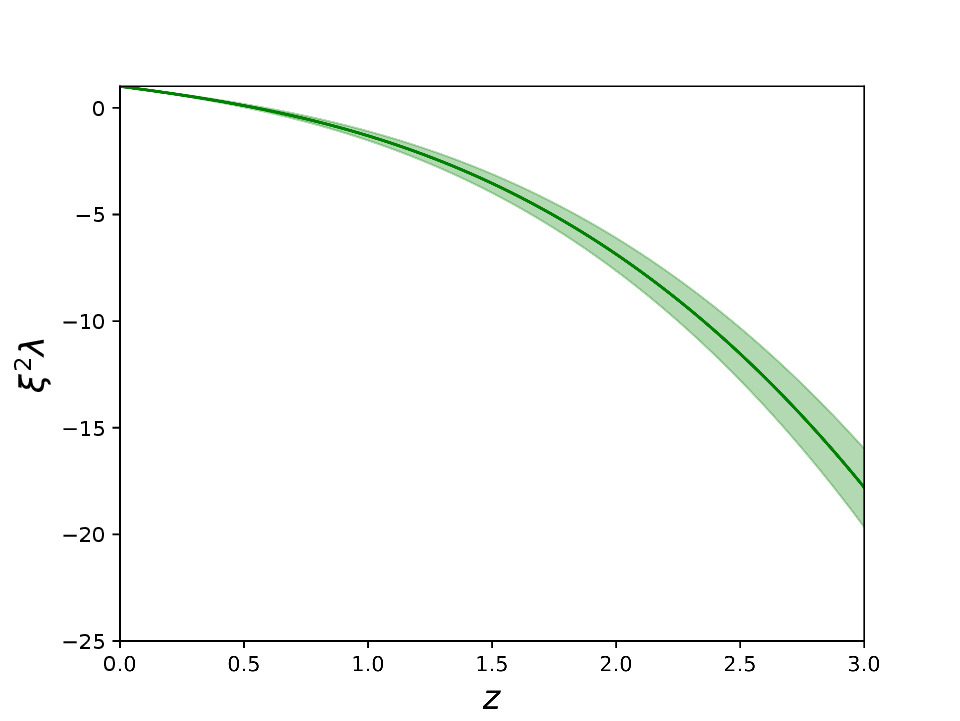}
	\caption{\label{figwefflambda} Variation of the parameter $w_{eff}(z)=P_{eff}(z)/r_{eff}(z)$ of the geometric equation of state of the dark energy (left panel), and  of the Lagrange multiplier $\lambda (z)$ (right panel) as a function of the redshift $z$ for the Mimetic-Weyl gravity model for the best fit values of the parameters as given by Eqs.~(\ref{bestfit}).The shaded area denotes the $1\sigma$ error. The dashed line represents the $\Lambda$CDM model.}
\end{figure*}

The redshift variations of the parameter $w_{eff}$ of the effective dark energy equation of state and of the Lagrange multiplier $\lambda$ are represented in Fig.~\ref{figwefflambda}. As one can see from the left panel of Fig.~\ref{figwefflambda}, $w_{eff}$ becomes singular at $z\approx 1.25$.  The presence of the singularity is due the existence of the zero of the effective geometric energy density at this redshift. Except in the neighborhood of the singular point, $w_{eff}$ takes values very close to $w_{eff}=-1$, and thus the model generates an effective cosmological constant on a large range of redshanks.  

The evolution of the Lagrange multiplier $\lambda$  is shown in the right panel of Fig.~\ref{figwefflambda}. In the considered redshift range the Lagrange multiplier takes only negative values, and it is a monotonically decreasing function of the redshift.

\begin{figure*}
	\includegraphics[scale=0.57]{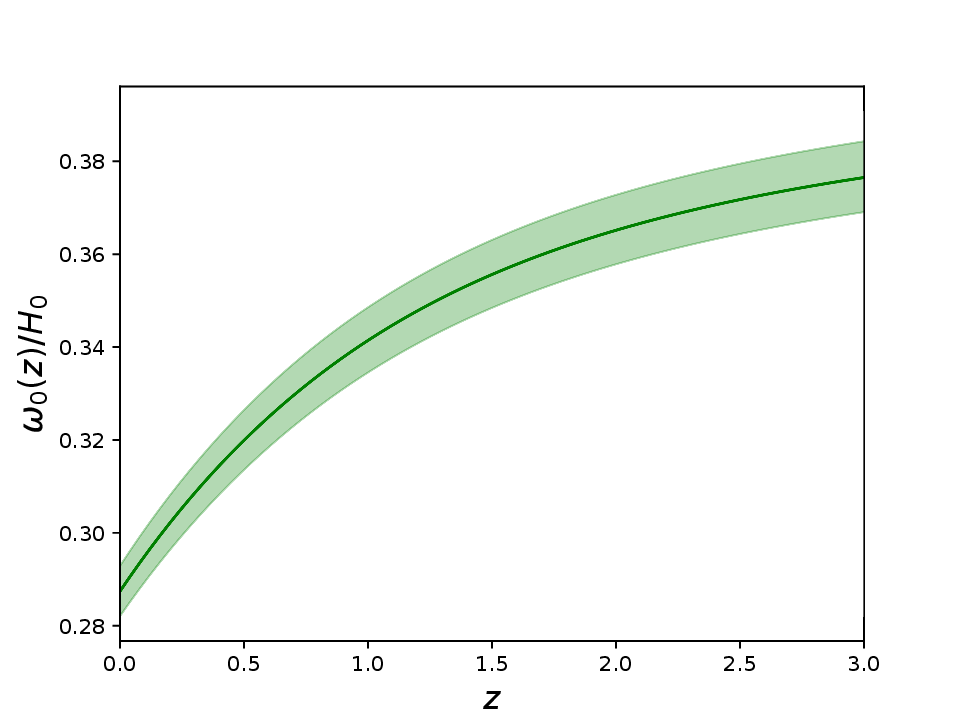}\includegraphics[scale=0.57]{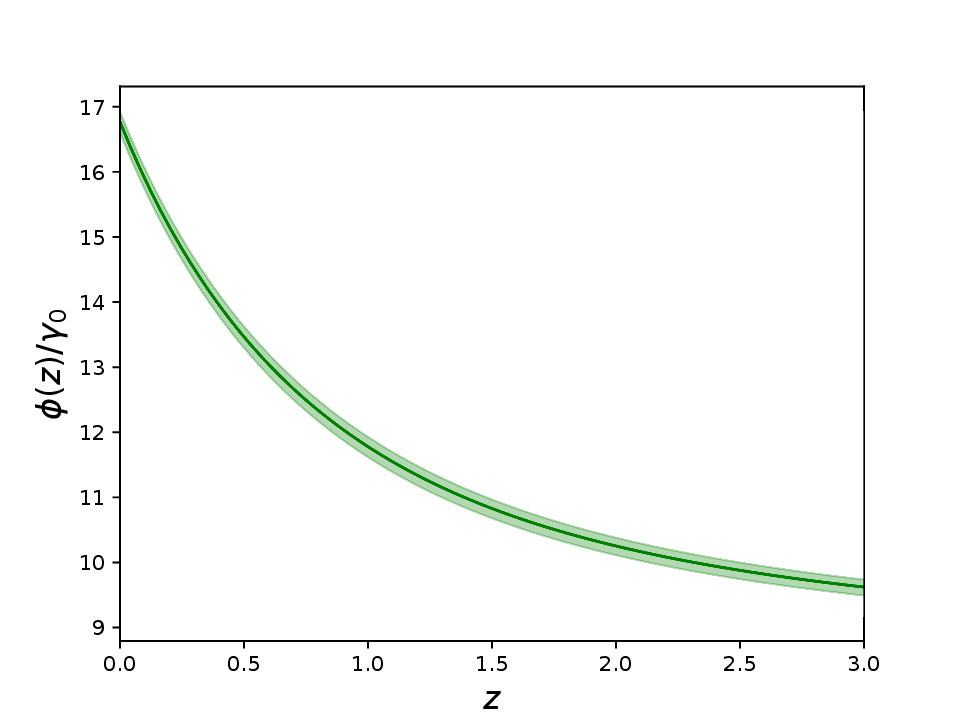}
	\caption{\label{figomegaphi} The behavior of the resealed Weyl vector component $\omega_0(z)$ (left panel) and of the scalar field $\Phi (z)$ (right panel) as a function of the redshift for the Mimetic Weyl gravity model for the best fit values of the parameters as given by Eqs.~(\ref{bestfit}). The shaded area denotes the $1\sigma$ error.}
\end{figure*}

The evolution of the Weyl vector $\omega_0$ and of the scalar field $\Phi$ can be seen, as functions of redshift,  in Fig.~\ref{figomegaphi}. The time component of the Weyl vector is a monotonically increasing function of the redshift (a monotonically decreasing function of time), and it has a nonlinear, almost quadratic, dependence on $z$. On the other hand the scalar field $\Phi$ is monotonically decreasing with the redshift, monotonically increasing in time. Both the Weyl vector and the scalar field take only positive values.

%\begin{figure}
%	\includegraphics[scale=0.57]{lambda-fitted.eps}%\includegraphics[scale=0.5]{tau-fitted.eps}
%	\caption{\label{figlambdatau} The behavior of the Lagrange multiplier $\lambda (z)$ (left panel) and of the dimensionless time coordinate $\tau (z)$ (right panel) as a %function of the redshift for the Mimetic-Weyl gravity model for the best fit values of the parameters as given by Eqs.~(\ref{bestfit}). The shaded area denotes the %$1\sigma$ error.}
%\end{figure}

\paragraph{$Om(z)$ diagnostic.} The $Om(z)$ diagnostic tool is an important theoretical method that allows to  distinguish alternative cosmological models from the standard  $\Lambda$CDM model. The $Om(z)$ diagnostic can be used to determine the nature of the considered dark energy model, and one could infer if the cosmological fluid  is a phantom-like fluid, a quintessence-like one, or it can be described by a simple cosmological constant.

The $Om(z)$ function is defined as \cite{69}
\begin{equation}\label{om}
Om(z)=\frac{\left( H(z)/H_0\right)^{2}-1}{(1+z)^{3}-1}.
\end{equation}

For the standard $\Lambda$CDM model, the function $Om(z)$ is a universal constant, and it is equal to the present day matter density parameter $\Omega_{m0}$. In the case of cosmological models with a constant parameter of the equation of state of dark energy, $w=p_{eff}/\rho_{eff}={\rm constant}$, a positive slope of $Om(z)$ indicates  a phantom behavior, while a negative slope points towards  a quintessence-like evolution.  For the standard  $\Lambda$CDM cosmology, and assuming that the dynamical dark energy fluid can be described by  a linear barotropic equation of state, with the parameter of EOS denoted by $w (z)$, the first Friedmann equation can be generally written as
\bea
\left(\frac{H(z)}{H 0}\right)^{2}&=&\Omega_{m 0}(1+z)^{3}+\left(1-\Omega_{m 0}\right) \nonumber\\
&&\times \operatorname{Exp}\left[3 \int_{0}^{z} \frac{1+w\left(z^{\prime}\right)}{1+z^{\prime}} d z^{\prime}\right].
\eea

For a constant $w$, we immediately find
\be
\left(\frac{H(z)}{H 0}\right)^{2}=\Omega_{m 0}(1+z)^{3}+\left(1-\Omega_{m 0}\right)(1+z)^{3(1+w)} ,
\ee
and
\be
Om(z)=\frac{\Omega_{m 0}(1+z)^{3}+\left(1-\Omega_{m 0}\right)(1+z)^{3(1+w)}-1}{(1+z)^{3}-1} .
\ee

In the case of the $\Lambda \mathrm{CDM}$ model, the parameter of the dark energy equation of state is $w=-1$, and thus
$Om(z)=\Omega_{m 0}$.

\begin{figure}[tbp]
\centering
	\includegraphics[scale=0.57]{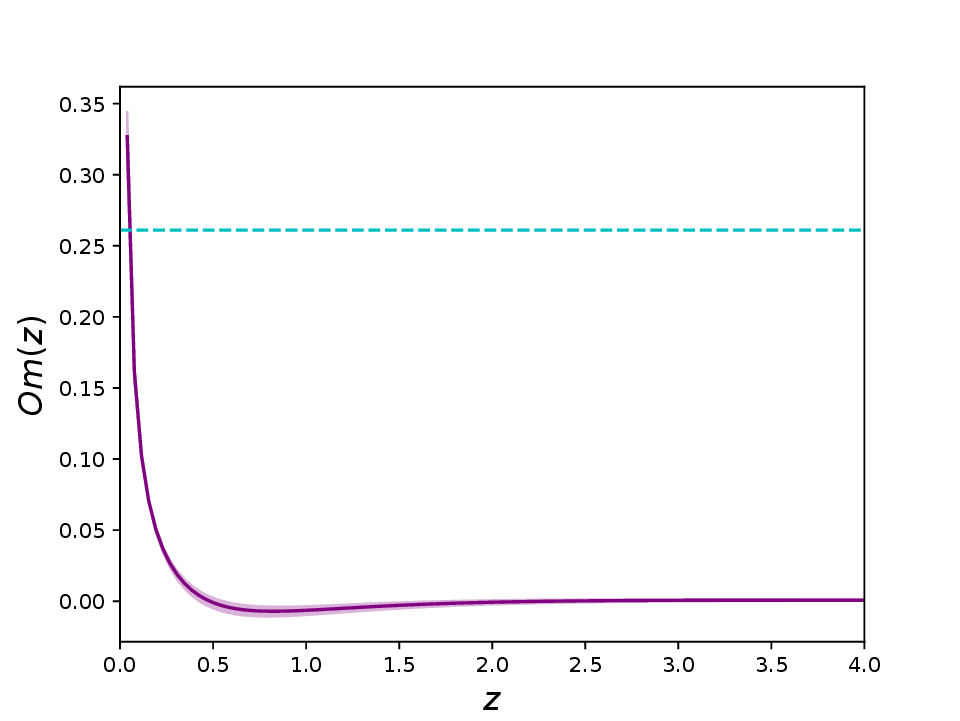}
	\caption{\label{figmatter} The redshift variation of the $Om(z)$ diagnostic function for the Mimetic Weyl geometric gravity model for the best fit values of the parameters as given by Eqs.~(\ref{bestfit}). The shaded area denotes the $1\sigma$ error. The dashed line represents the $\Lambda$CDM model.}
\end{figure}

The variation with respect to the redshift of the $Om(z)$ diagnostic function for the Mimetic Weyl geometric gravity cosmological model is represented in Fig.~\ref{figmatter}. In the redshift range $0\leq z\leq 0.7$, the slope of the $Om(z)$ function is negative, indicating a quintessence-like behavior of the cosmological fluid. For $z>0.7$, the slope of $Om(z)$ becomes positive, indicating a transition from the quintessence-like behavior to a phantom like behavior.

\section{Discussions and final remarks}\label{sect3}

In the present study, we have explored an extension of the Weyl geometric gravity theory that incorporates the mimetic properties of the metric, which have been extensively used to investigate dark matter properties, and cosmological evolution. We adopted the Weyl conformal geometry as the foundational framework for exploring the gravitational implications of the nonmetricity, conformal transformation and invariance, and of the scalar representation of the metric. Introducing the concept of nonmetricity requires
to redefine the compatibility condition between the metric tensor and the connection, by allowing that the divergence of the metric tensor does not vanish. We have also adopted, as a starting point, the simplest possible conformally invariant action, which must be quadratic in the Weyl scalar $\tilde{R}$, and can also include the strength of the Weyl vector, plus the matter contribution.

The first essential step in building our model is the linearization of the action in terms of the Weyl scalar, via the introduction of a scalar field, so that $\tilde{R}^2=-2\phi^2\tilde{R}-\phi^4$. With the help of this transformation one obtains a scalar-vector-tensor action, which is constructed from the Weyl scalar (in a linear representation), the Weyl, vector, and the scalar field $\phi$.

Furthermore, we have considered a modification to the basic Weyl geometric theory that involves the representation of the physical metric in terms of an auxiliary metric, obtained through a conformal transformation that depends on a scalar field. This mimetic type representation of the metric has been extensively investigated in many previous studies, and it was initially proposed as a modified gravity explanation of the dark matter properties \cite{M1} in terms of
geometrical characteristics, and it is essentially based on the representation of the physical metric in terms of an auxiliary, scalar field dependent metric.

In our approach we have assumed that the mimetic scalar field is identical to the scalar field of the Weyl geometric gravity, and which can be represented in purely Weylian geometric terms, by taking into account that 
\be
\phi=\sqrt{-\tilde{R}}=\sqrt{-R+3\alpha \nabla_\mu \omega ^\mu+\left(\frac{3\alpha ^2}{2}\right)\omega ^2}. 
\ee
Thus for the mimetic conformal transformation of the metric we obtain
\begin{widetext}
\begin{eqnarray}
g_{\mu \nu } =\tilde{g}^{\sigma \lambda }\frac{\nabla _{\sigma }\tilde{R}%
\nabla _{\lambda }\tilde{R}}{4\tilde{R}}\tilde{g}_{\mu \nu }
=\tilde{g}^{\sigma \lambda }\frac{\left( -\nabla _{\sigma }R+3\alpha
\nabla _{\sigma }J+3\alpha ^{2}\nabla _{\sigma }\omega ^{2}\right) \left(
-\nabla _{\lambda }R+3\alpha \nabla _{\lambda }J+3\alpha ^{2}\nabla
_{\lambda }\omega ^{2}\right) }{\left( -R+3\alpha J+\frac{3}{2}\alpha
^{2}\omega ^{2}\right) }\tilde{g}_{\mu \nu },
\end{eqnarray}
where we have denoted $J=\nabla _\mu \omega ^\mu$.
\end{widetext}

Thus, in the present Weylian extension of the Riemannian geometrical gravitational theories, we have implemented the mimetic field approach without introducing a new additional scalar field degree of freedom. Hence, the entire mathematical formalism is dependent on the Weyl vector, and the auxiliary scalar field, which can also be interpreted as a mimetic field. However, in the present approach the scalar field is of purely geometric origin, since it is directly related to the geometry of the Weyl, and by extension, of the Riemann manifold.

In the initial formulation of the present theory, which did begin with writing down the simple quadratic gravitational Lagrangian density, followed by its linearization,  an important problem is the introduction of an effective matter term. Note that adding the simple matter Lagrangian breaks the conformal invariance of the action, and thus one must resort to a more general matter term.   The effective matter Lagrangian density $\mathcal{L}_m$ was assumed  to be an arbitrary function of the ordinary matter
Lagrangian $L_m$, the square of the Weyl vector field $\omega ^2$, and of the scalar field $\phi$, so that $\mathcal{L}_m=\mathcal{L}_m\left(L_m, \omega ^2,\phi\right)$.  It should be noted that  both the geometric sector and the matter sector of the total action must remain invariant under conformal transformations. However, in the case of matter, it is not the action, but its variation that must preserve the conformal invariance.  The mimetic field was added with the help of the Lagrange multiplier. From the proposed action and the variational principle, we have derived the gravitational field equations by varying the action with respect to the metric tensor. The constraint on the mimetic field was also obtained after varying the action with respect to the Lagrange multiplier. The equations thus obtained represent a generalization of the equations of the Weyl geometric gravity theory \cite{W9}.

Mimetic gravity has also the potential of describing not only dark matter, as initially proposed in \cite{M1}, but also inflation \cite{V0,V1,V2,V3}, with the mimetic matter playing the role of the inflaton. Early accelerating, or at least marginally accelerating phases can also be obtained in the mimetic Weyl geometric theory. In the limit of small times, $\tau \rightarrow 0$,  by assuming $\alpha ^2>> \beta /\tau ^2$,  Eq.~(\ref{e:d1111}) can be approximated as 
\be\label{eqn}
2\frac{HD}{d\tau}+3h^2-\frac{1}{\tau^2}=\sigma _0.
\ee   
For $\sigma _0\approx 0$, Eq.~(\ref{eqn}) has the solution 
\be
h(\tau)=\frac{1}{t}-\frac{4c_1}{3\tau\left(\tau ^2+c_1\right)},
\ee
where $c_1$ is an arbitrary constant of integration. For the scale factor of the early Universe we obtain
\be
a(\tau)=\frac{\left(\tau ^2+c_1\right)^{2/3}}{\tau ^{1/3}},
\ee 
with the deceleration parameter given by
\be
q(\tau)=-1+\frac{9\tau ^4-18c_1\tau ^2-3c_1^2}{\left(3\tau ^2-c_1\right)^2}.
\ee
For $c_1=0$, or $\tau ^2>>c_1$, we obtain $a(\tau)=\tau$, and $q(\tau)=0$. In this case the Universe begins its evolution in a marginally accelerating phase.  For a moment in time when $\tau =\sqrt{c_1}\sqrt{1+ 2/\sqrt{3}}$, the Universe is in a de Sitter type expansionary phase. Hence, mimetic Weyl geometric gravity has also the potential of describing the very early stages of the cosmic evolution.

One of the important tests of the gravitational theories is their consistency with respect to the absence of ghost instabilities. The initial mimetic model \cite{M1}  describes
a regular pressureless dust, with the energy density of the mimetic matter remaining positive during the temporal evolution. But for certain types of configurations, the theory may become unstable \cite{Ch}. y. In general, the formulation of the mimetic theory by using the physical metric $g_{\mu \nu}^{phys}$ could lead to a theory containing higher order derivatives of the scalar field $\phi$, which could lead to the emergence of ghosts \cite{b4}. However, as shown in \cite{Ch}, alternative modified theories of gravity, which are based on a vector field (tensor-vector theory) or a vector field and a scalar field (scalar-vector-tensor theory), are free of ghost instabilities. The present mimetic Weyl geometric gravity belongs to the class of the scalar-vector-tensor theories, similar to those considered in \cite{Ch}, and therefore the results on the existence of instabilities obtained in \cite{Ch} can be extended to the case of the present theory. Moreover, in the theory no higher order derivatives of the scalar field do appear. The above mentioned results may suggest that the mimetic Weyl geometric theory is free of ghosts.   However, before reaching a definitive conclusion on the problem of ghost instabilities a detailed Hamiltonian analysis of the model is necessary.

We have studied the cosmological implications of the Mimetic Weyl geometric gravity theory by considering the case of the isotropic, flat, and homogeneous  FLRW Universe, by analysing the underlying evolution of the corresponding cosmological model. The expression of the energy-momentum tensor was chosen in the form of a perfect fluid. To maintain the isotropy and homogeneity of the Universe, only the presence of the time component of the Weyl vector is allowed in the mathematical formalism, and all physical and geometrical quantities are time dependent only. Furthermore, the constraint on the mimetic field can be solved immediately, and it gives a simple expression for the scalar field as being proportional to $t^2$, $\phi (t)\propto t^2$. The Weyl vector can also be expressed in terms of $\phi$.

The generalized Friedmann equations for the Mimetic Weyl geometric gravity theory were derived in a general form, and their dimensionless representation was also obtained. For a direct validation of the model with the observational data we have also formulated the generalized Friedmann equations in the redshift representation.
The solution of the system of cosmological equations can only be determined through numerical methods. The model contains
six  parameters, whose optimal values were  obtained after constraining the model with a small sample of observational values of the Hubble function.

The best-fit values for the model  parameters were determined by performing a Likelihood analysis at the 1$\sigma$ confidence level, as shown in Eqs~(\ref{68}) and (\ref{69}). After obtaining the best fit values of the model parameters, we have conducted an in-depth comparative analysis between our theoretical model, the $\Lambda$CDM model, based on Einstein’s gravity, and the observational data. The purpose of this analysis is to provide a comprehensive evaluation of the cosmological implications of the model, by comparing it with the Hubble data points, and with the standard cosmological paradigm.

The FLRW cosmological  model based on the Mimetic Weyl geometric gravity theory  accurately describes the Hubble data points through the normalized Hubble function. The result for the present day value of the Hubble function, $H_0 = 71.476^{+0.095}_{-0.091}$ km/s/Mpc  from the best-fit analysis of the observational data by the Mimetic Weyl geometric model aligns relatively well with the SH0ES value $H_0=73.30$ km/s/Mpc, but still predicts a somehow lower value, thus slightly alleviating  the existing tension surrounding the Hubble constant.

The dynamical properties of the Universe are then accurately determined through the analysis of the cosmographic quantities, the deceleration $q$, the jerk $j$, and the snap $s$ parameters, respectively. In the Mimetic Weyl geometric gravity model, $q$ takes on higher values for redshanks $z > 1$, indicating a less accelerating expansion than the one predicted in $\Lambda$CDM. The jerk parameter converges to the standard $\Lambda$CDM model value at higher redshanks, while the snap parameter deviates significantly in all the considered redshift range from the standard cosmological value.

When considering the matter density, we observe a good concordance  with the $\Lambda$CDM behavior. The resealed Weyl vector field gradually increases across the redshift
space, while the scalar field decreases. The evolution of the Lagrange multiplier shows a gradual decrease, with the variable taking negative values at higher redshanks. The dimensionless time coordinate has a nonlinear dependence on the redshift variable. There is also  a transition of the effective dark energy from the quintessence state to the phantom regime. Hence one can assume the existence of a two phase dynamical DE model. When correlating this with the mimetic scalar field $\phi$, we find that for a larger value of $\phi$, we have a quintessence-like dark energy, while a decrease in the scalar field leads to a shift to the phantom phase, with the energy density increases with the redshift.

Even if the present model can give a good description of the cosmological data, several other tests are necessary to test its relevance for the understanding of the gravitational interaction. In particular, the astrophysical and Solar System tests of the mimetic Weyl geometric gravity could lead to a detailed investigation of the physical effects of the mimetic scalar field at different length scales. Black hole solutions in mimetic gravity were considered in \cite{N1}. Obtaining a black hole solution in the standard approach to mimetic theory is not a trivial task, since to obtain the solution one needs both timelike and spacelike vectors $\partial ^\mu \phi$. Schwarzschild type solutions are possible, but with the mimetic field propagating as a scalar hair (stealth Schwarzschild solutions), and for the choice of $\lambda =0$ \cite{N1}. On the other hand, for $\lambda \neq 0$, no solutions with a horizon can be found, and the obtained solutions have a singularity only at the center $r=0$, and at the branch cut off $r=r_f$. These solutions correspond to naked singularities \cite{N1}. However, a mimetic black hole solution can be constructed by gluing the exterior static spherically symmetric solution to a time-dependent anisotropic geometry, which describes the interior of the black hole. 

However, in a detailed comparison of the black hole and naked singularity solutions obtained in \cite{N1}, in \cite{Va1} it was pointed out that the shadow properties of these mimetic gravity objects are pathological. Thus,  the black hole casts a shadow that is too small, while the naked singularity does not cast a shadow. Thus, one can conclude that the Event Horizon Telescope images of M87* and Sgr A* represent a severe challenge to the black hole solutions of standard mimetic gravity, by ruling out the baseline version of the theory. Therefore, these observational results has cast doubt on the potential of mimetic gravity as an alternative to the dark matter paradigm.

One possibility to solve this problem is to modify the form of the Lagrange multiplier constraint \cite{Va2}.  By using the new approach,  the cosmological implications  and the black hole solutions in mimetic gravity with scalar field potential, and in the scalar mimetic $f(R)$ gravity were investigated in \cite{Va3}.  Two black hole solutions, also containing the Schwarzschild and Hayward geometries, have been found. The shadow and the radius of the photon sphere for these black holes have been also obtained, and by comparing the black holes shadow radii with the observational bounds from M87* and Sgr A* it was shown that they are consistent with the observational data.

The role of the scalar field potential is also essential in obtaining  black hole solutions in modified gravity theories, like, for example, in $f(G)$ theory. In \cite{Va4} spherically symmetric solutions of the ghost-free $f(G)$ gravity have been obtained, with the solutions including the Reissner-Nordstr\"{o}m blac or the Hayward black holes. A black hole that  contains the Arnowitt-Deser-Misner (ADM) mass, the horizon radius, and the radius of the photon sphere as independent parameters is also constructed.  

 Black hole solutions in the Weyl geometric gravity have been studied extensively in \cite{W3}. An exact analytical black hole solution that generalizes the Schwarzschild one can be obtained by assuming that the Weyl vector has only a radial component. For other choices of the functional form of the Weyl vector black hole solutions can be obtained only numerically, with the numerical investigations indicating the formation of an event horizon. The solutions found in \cite{W3} can, at least in principle, be generalized to the case of the mimetic Weyl geometric gravity. A significant difference with respect to the standard general relativistic case may be represented by the fact that the solutions that do not require the presence of the timelike vector $\partial ^\mu \phi$ could also be obtained. However, solutions describing naked singularities may also be found.  It is important to point out that a successful theory must not only describe 
inflation in the early Universe, or late time cosmic acceleration, but it must also satisfy the observational constraints coming from the astrophysical observations of the black holes, or from the Solar System tests.            

The present study conclusively demonstrated that the Mimetic Weyl geometric gravity theory allows  accurate theoretical predictions that are consistent  with the observational data for the  Hubble function,  as can also be seen in the values of the  $\chi ^2$ function of the fit. Hence, the Mimetic Weyl geometric gravity theory
can provide an alternative explanation for the accelerated expansion of the Universe, without resorting to the mysterious cosmological constant. However, in order
to get a more comprehensive understanding and a consistent validation of the theory, further data points at higher redshanks for the Hubble function are needed, and other observational datasets must be analyzed in their full generality.

%\clearpage
%\newpage
%\mbox{~}

\section*{Acknowledgments}

We would like to thank the two anonymous reviewers for comments and suggestions that helped us to improve our manuscript.

\end{document}